\documentstyle{mn}

\input{epsf}
\def\eps@scaling{.95}
\def\plotone#1{\centering \leavevmode
\epsfxsize=\eps@scaling\columnwidth \epsfbox{#1}}

\title{Electron-Cyclotron-Maser Observable Modes}
\author[A. Stupp]{A. Stupp \\
School of Physics and Astronomy, Tel-Aviv University, Israel}

\begin{document}

\maketitle
\begin{abstract}
  We investigated wave amplification through the 
Electron-Cyclotron-Maser mechanism.  \par
   We calculated absorption and emission coefficients 
without any approximations, also taking into account absorption by the 
ambient thermal plasma. A power-law energy distribution for the fast electrons was used, 
as is indicated by X-ray and microwave observations.  \par
  We developed a model for the saturation length and amplification ratio of the maser,
scanned a large parameter space, and calculated the absorption
and emission coefficients for every frequency and angle. \par
  Previous studies concluded that the unobservable Z-mode
dominates in the  $\nu_{\rm p} \approx \nu_{\rm B}$ region, and that 
millisecond spikes are produced in the region $\nu_{\rm p}/\nu_{\rm B}<0.25$. 
We find that the observable O-mode and X-mode can produce emission
in the $0.8<\nu_{\rm p}/\nu_{\rm B}<2$ region,  which is expected at the footpoints 
of a flaring magnetic loop. \par
 The important criterion for observability is the saturation 
length and not the growth rate, as was assumed previously, and even 
when the Z-mode is the most strongly amplified, less strongly amplified 
O-mode or X-mode are still intense enough to be observed.  \par
   The brightness temperature computed with our model for the saturation length
is found to be of order $10^{16}$ K and higher. The emission is usually 
at a frequency of $2.06\nu_{\rm B}$, and at angles 30--60 degrees to 
the magnetic field. The rise time of the amplified emission to maximum is a 
few tenths of a millisecond to a few milliseconds, and the emission persists 
for as long as new fast electrons arrive into the maser region.
\end{abstract}


\section{Introduction}
  Electromagnetic emission generated by relativistic electrons moving 
in a magnetic field is referred to as {\it synchrotron} radiation if the 
electrons are ultra-relativistic, and as {\it gyrosynchrotron} if the 
electrons are mildly relativistic. In the present paper we deal only 
with gyrosynchrotron radiation, using the formalism developed by Ramaty
\shortcite{R}, except that we also include the longitudinal 
component of the radiation which is important at low multiples 
of the cyclotron frequency. We also take into account the correction 
pointed out by Trulsen and Fejer \shortcite{Trulsen}. This formalism has been 
tested by other workers (e.g. Holman \& Benka 1992) and was found to 
be accurate. Gyrosynchrotron absorption can become negative and 
this amplification process is referred to as the Electron-Cyclotron-Maser
(ECM). \par
  The observation of millisecond microwave spikes from the Sun (reviewed 
by Benz 1986) has been interpreted as gyrosynchrotron maser 
emission since the late 1970`s (Wu \& Lee 1979; Holman, Eichler \& Kundu 1980), 
and it is now the most widely accepted explanation for the 
spike emission \cite{BenzBook}.  Spikes have been observed in the 
frequency range 0.3--8 GHz, exhibit a duration of a few 
milliseconds to a hundred milliseconds, a narrow bandwidth 
(0.5--2 per cent or a few MHz), and until recently have not been resolved in 
angle \cite{Benzetal}.  By assuming
that the homogeneity scale of the magnetic field is of order $10^7$ cm, and
that this is the size of the source, it is possible to deduce a characteristic 
brightness temperature of $10^{15-18}$ K . \par
   In order that ECM appears the electron number distribution must either increase
towards higher energies or an anisotropy in pitch angle must exist, 
as in a loss cone distribution. 
  A loss cone forms naturally where the magnetic field increases, as
electrons with small pitch angles continue and electrons with
large pitch angles mirror. At the footpoints of flare loops the density
becomes high enough that electrons continuing into the chromosphere lose
their energy and do not reflect back upwards, and thus a loss-cone
distribution appears (Aschwanden \& Benz 1988; 
Kuncic \& Robinson 1993). 
We therefore expect maser emission from the
region of the footpoints. \par
   The study of the ECM mechanism is of interest because of the radio emission, which
serves as a probe into conditions in the flare, and also because 
of the energetic particle precipitation which produces X-rays, and for local heating of 
coronal material.
  The ECM reduces the trapping time and increases the 
flux of precipitating particles by causing a reduction in the pitch angle 
of electrons within a short distance \cite{BenzBook}. 
Electrons which otherwise would be mirrored back and forth are thus precipitated 
into the chromosphere after one reflection. The maser thus transfers energy from the 
fast particle distribution to the ambient  plama, and serves as a heating mechanism.  \par
  Several papers have explored the possibilities of ECM. 
Melrose and Dulk \shortcite{D+M} have used a semi-relativistic approximation to derive
approximate formulae for the frequency of the largest growth rate, and for the
growth rate. Winglee \shortcite{Winglee} approximated the effects of the ambient plasma 
temperature on the maser emission. Aschwanden (Aschwanden 1990a,b) followed the 
diffusion of electrons into the loss-cone as a result of the maser emission, and 
therefore the self-consistent closure of the loss-cone as kinetic energy is 
transformed to radiation through the maser. Aschwanden \shortcite{Aschwanden b} 
concluded that the millisecond spikes are produced in the region of 
$\nu_{\rm p}/\nu_{\rm B}<0.25$.
Kuncic and Robinson \shortcite{Kuncic}  performed ray-tracing in a model loop 
with dipole field,
and concluded that maser emission can escape from lower levels.
Fleishman and Yastrebov \shortcite{Fleishman1} present results which show 
simultaneous amplification in two or more frequencies. \par
   An important difference in our work is that   
all previous studies concentrated on the {\it growth rate}
of the amplified wave, and defined the {\it dominant mode} as the particular
wave -- frequency $\nu$ and direction to the magnetic field $\theta$ -- with the 
largest growth rate. However, the important criterion for long-term emission
is the {\it saturation length}, which is the distance beyond which the maser
ceases to amplify, and not the growth rate \cite{BenzBook}. 
We compute the absorption coefficient and, using our model, the 
saturation length, and determine which wave will be observed, and at what 
intensity. \par
  Another difference is that most previous studies used soft energy distributions for the fast 
electrons, such as the Dory, Guest, Harris distribution \cite{Winglee2} or 
a hot thermal component distribution (Aschwanden 1990a,b). However,  observations 
of Microwaves (MW) and X-rays from solar flares indicate a power law energy
spectrum with energy up to 10 MeV and even higher, and with a power index
typically $\delta=3$, and in the range $\delta=2-6$ (Ramaty 1969; Zirin 1989;
Fleishman \& Yastrebov 1994b for other references). 
We therefore chose to use a power law distribution function for our fast 
non-thermal electrons.
We also use an idealized one sided loss-cone distribution \cite{D+M} 
while other studies use a one sided $sin^N(\phi)$ distribution 
(Aschwanden 1990a,b) or a two-sided gaussian 
(Fleishman \& Yastrebov 1994a,b) pitch angle distribution. \par
  The ratio $\nu_{\rm p}/\nu_{\rm B}$, where $\nu_{\rm p}=\sqrt{n_e/m/\pi}\ e$ is the plasma 
frequency and $\nu_{\rm B}=eB/(m\gamma c)/(2\pi)$ the cyclotron frequency, 
is an important parameter for ECM emission. It is used to distinguish regions
where different magneto-ionic modes dominate.   
From the previous studies we conclude that the dominant mode 
depends on the distribution function of the fast electrons.
In some studies the fundamental eXtraOrdinary (XO) mode is dominant 
for $\nu_{\rm p}/ \nu_{\rm B} < 0.3$ \cite{B+A2}, while in others
the fundamental Ordinary Mode (OM) is dominant for $\nu_{\rm p}/ \nu_{\rm B} < 0.5$ 
\cite{Winglee2}.  In some the fundamental OM is
dominant for $0.3 < \nu_{\rm p}/\nu_{\rm B}<1$ \cite{B+A2}, and in others
the second harmonic XO is dominant for $0.5 < \nu_{\rm p}/\nu_{\rm B}<1.5$ 
\cite{Winglee2}.
The  unobservable Z-Mode (ZM) is dominant for $\nu_{\rm p}/\nu_{\rm B} > 1.5$ 
\cite{Winglee2}, or even for $\nu_{\rm p}/\nu_{\rm B} > 0.5$ \cite{Fleishman2}. 
     In our work the ZM is the most highly amplified mode (i.e. with the
largest in absolute value negative absorption) for the range 
$0.5 < \nu_{\rm p}/\nu_{\rm B} < 1.1$ and for $1.8 < \nu_{\rm p}/\nu_{\rm B}$,
similar to what was found by others who have used a power 
law \cite{Fleishman2}.  \par
  The high amplification of the ZM would seem to indicate that this unobservable
mode is the dominant mode, and would therefore preclude the maser mechanism 
as the producer of MW spikes, except in the narrow range $1.1 < \nu_{\rm p}/\nu_{\rm B}<1.8$. 
  We show that for the range $\nu_{\rm p}/ \nu_{\rm B} < 2$ it is possible for the OM 
or the XO mode to be amplified enough that they become observable. 
Our absorption-coefficient approach allows a simple procedure to estimate 
the saturation length, which is difficult to derive from growth rate calculations,
and is the important parameter in deciding which mode
dominates \cite{BenzBook}. 
Our derived saturation lengths are consistent
with the assumed homogeneity of the magnetic field and density. We
therefore conclude that the Electron-Cyclotron-Maser mechanism
can produce microwave spikes. \par
  In section 2  we give the theory and equations for gyrosynchrotron
emission and wave propagation through cold-plasma.
  In section 3 we show that the gyrosynchrotron is the important
process in absorption as well as emission, for the frequencies and
plasma parameters we assume.
  Section 4   describes our estimate of the energy available to the
loss-cone, and a derivation of the saturation length is given.
  Section 5 describes our calculations and the parameters we used.
  In section 6  we discuss the observability of the maser emission under
different conditions, in section 7 we give a comparison with previous studies,
and in section 8 we summarize our results.  

\section{Gyrosynchrotron Absorption}
  Solar flares are known to be a complex phenomena. The solar flare plasma is 
composed of both an ambient thermal plasma,  and a population of fast particles. 
It is customary to simplify and assume that any waves which 
are excited are a phenomena of the ambient plasma alone. The properties 
of the ambient plasma therefore determine which waves are excited and 
how the waves propagate. However, the fast particles can absorb waves 
present in the ambient plasma, and also emit in these waves` frequencies. 
  We used the cold plasma approximation to describe the ambient plasma,
and assumed the waves which result from  magnetoionic theory
\cite{MelroseB}. 
    For a given electron distribution $f({\bmath p})$ and wave $(\nu,\theta)$
it is then possible to compute  the emission and absorption coefficients, which describe 
the interaction of the wave with the distribution. \par
  The cold plasma parameters which determine the propagation are the refraction index
and the components of  the polarization vector \cite{MelroseB}:
\begin{eqnarray}
\label{polvec}
&&\hat\epsilon_\sigma = { {L_{\sigma}({\bmath k}){\hat k} +
T_\sigma({\bmath k}) {\bmath t} + i{\bmath c}} \over 
\sqrt{1+L_\sigma^2+T_\sigma^2} }
\end{eqnarray}

where  $\bmath k$ is the wave vector,$\hat k$ a unit vector in the 
$\bmath k$ direciton, $\bmath c$ a unit vector perpendicular to $\bmath k$ and to 
the magnetic field ${\bmath B}$, and ${\bmath t}$ is a unit vector perpendicular to 
${\bf k}$, in the plane of ${\bf k}$ and ${\bf B}$. 
Also $T_\sigma$ and $L_\sigma$ are functions depending on
$\bmath k$, and $\sigma$ is  `+` (OM) or `-` (XO mode).
The transverse component of the polarization is given by $T_\sigma$ and the 
longitudinal component is described by $L_\sigma$.
We assumed ${\bmath B}$ to be in the $\hat z$ direction, and ${\bmath k}$ to be
in the $(x,z)$ plane with angle $\theta$ to the magnetic field. With
these definitions ${\bmath c}=\hat y$ , $\hat k=(\sin(\theta),0,\cos(\theta))$,
and ${\bmath t}=(\cos(\theta),0,-\sin(\theta))$. \par
  The cold plasma modes are the OM, which exists for 
\begin{equation}
\nu>\nu_{\rm p}
\end{equation}
the XO which exists for 
\begin{equation}
\nu>\nu_{\rm x}=\nu_{\rm B}/2 + \sqrt{\nu_{\rm p}^2+\nu_{\rm B}^2/4}
\end{equation}
and the low frequency branch of the XO, called here the ZM, which
exists for 
\begin{eqnarray}
\nonumber
\nu_{\rm x}-\nu_{\rm B} & < &\nu < \sqrt{0.5} \times \\
&&\nonumber
 \sqrt{  (\nu_{\rm p}^2+\nu_{\rm B}^2)+
[(\nu_{\rm p}^2+\nu_{\rm B}^2)^2-4\nu_{\rm p}^2\nu_{\rm B}^2\cos^2(\theta)]^{0.5} }
\end{eqnarray}
and is bound from above by the upper-hybrid frequency 
\begin{equation}
\nonumber
\nu_{\rm uh}=\sqrt{\nu_{\rm B}^2+\nu_{\rm p}^2}
\end{equation}
The Whistler mode exists for 
\begin{equation}
\nonumber
\nu< {\sqrt{ { (\nu_{\rm p}^2+\nu_{\rm B}^2)-
[(\nu_{\rm p}^2+\nu_{\rm B}^2)^2-4\nu_{\rm p}^2\nu_{\rm B}^2\cos^2(\theta)]^{0.5}}\over{2} } }
\end{equation}
and is a low frequency mode which we ignore \cite{MelroseB}. \par
 The equations for the cold-plasma parameters are (Ramaty 1969;
Melrose 1989)
\begin{eqnarray}
\label{Tpm}
T_\pm &=&
[2\nu (\nu_{\rm p}^2-\nu^2)\cos(\theta)] \times \\
&&\nonumber
\big[ -\nu^2\nu_{\rm B} \sin^2(\theta) \pm \\
&&\nonumber
\sqrt{ \nu^4\nu_{\rm B}^2\sin^4(\theta)+ 
4 \nu^2(\nu_{\rm p}^2-\nu^2)^2\cos^2(\theta) }\ \big] ^{-1}
\end{eqnarray}

\begin{equation}
\label{Lpm}
L_\pm={{\nu_{\rm p}^2\nu_{\rm B}\sin(\theta)}\over{\nu^3-\nu\nu_{\rm p}^2}}
{{\nu T_\pm}\over{\nu T_\pm-\nu_{\rm B}\cos(\theta)}}
\end{equation}

And the refraction index $n_{\rm ref}$, also referred to as $n_\pm$, is given by
\begin{eqnarray}
n_\pm^2 &=&  1 + [2\nu_{\rm p}^2(\nu_{\rm p}^2-\nu^2)] \times \\
&&\nonumber
\big[ -2\nu^2(\nu_{\rm p}^2-\nu^2)-\nu^2\nu_{\rm B}^2\sin^2(\theta) \pm \\
&&\nonumber
\sqrt{\nu^4\nu_{\rm B}^4\sin^4(\theta)+
4\nu^2\nu_{\rm B}^2(\nu_{\rm p}^2-\nu^2)^2\cos^2(\theta)}\ \big] ^{-1}
\end{eqnarray}

With the symbols being $\nu_{\rm p}$ the plasma frequency, $\nu_{\rm B}$ the cyclotron
frequency, and $\nu$ the emitted frequency. \par
   The cold plasma approximation is valid only if the two following
criteria are are true \cite{MelroseB}
\begin{equation}
\label{criteria1}
\left({{\nu- s\nu_{\rm B}}\over{n_{\rm ref}\nu\cos(\theta)}}\right)^2  >> 
{{k_{\rm B} T}\over{m_e c^2}} = 0.001683 {T\over{10^7\ {\rm K}}}
\end{equation}

and

\begin{equation}
\label{criteria2}
\left({{\nu_{\rm B}}\over{n_{\rm ref}\nu\sin(\theta)}}\right)^2  >> 
{{k_{\rm B} T}\over{m_e c^2}}= 0.001683 {\rm T\over{10^7\ K}}
\end{equation}
  Where we have used $k_{||}=n_{\rm ref}\omega\cos(\theta)/c$ and 
$v_{\rm thermal}=\sqrt{k_{\rm B} T /m_e}$  to get the above from Melrose`s equations. 
The $s$ multiplier in the first criterion is an integer such that $s\nu_{\rm B}$ is 
the closest cyclotron harmonic to $\nu$. \par
  Our rule for when the cold plasma approximation ceases to be valid is when
the left side of the criteria is less then ten times larger than the right side. 
The cold plasma approximation fails near the resonances of the cold plasma 
modes at the cyclotron harmonics, and for the low-frequency  Z-mode 
when the frequency of the waves is near the upper-hybrid frequency. \par
 Most authors calculate the growth rate, but our derivation follows
Ramaty`s \shortcite{R} in calculating the absorption coefficient 
which is (Ramaty 1969; Aschwanden 1990a):

\begin{eqnarray}
\label{gyroe}
k_\pm \left( {\nu ,\theta } \right)& = &
{1 \over B}4\pi ^2e
{{2\pi } \over {\left| \cos \left( \theta  \right) \right|}}
{{\nu_{\rm B}} \over \nu }{1 \over {n_\pm ^2}}
\int_1^\infty  {\rm d}\gamma {{u\left( \gamma  \right)} \over \beta }\times \\
&&\nonumber
\sum\limits_{\rm s=s_{\rm 1}}^{\rm s_{\rm 2}}
{{g\left( {\phi_{\rm s}} \right)} \over {1+T_\pm ^2}}
\Bigl[ \beta \sin \left( {\phi_{\rm s}} \right)J_{\rm s}^\prime\left( {x_{\rm s}} \right)+ 
{{L_\pm}\over{n_\pm}}J_{\rm s}\left(x_{\rm s} \right)+ \\
&&\nonumber
T_\pm \left({{\cot \left( \theta  \right)} \over {n_\pm }}-
{{\beta \cos \left( {\phi_{\rm s}} \right)} \over {\sin \left( \theta \right)}}
 \right)J_{\rm s}\left( {x_{\rm s}} \right) \Bigr]^2\times\\
&&\nonumber
\Bigl[ {{-\beta \gamma ^2} \over {u\left( \gamma  \right)}}      
{{\rm d} \over {{\rm d}\gamma }}
{{u\left( \gamma  \right)} \over {\beta \gamma ^2}}+ \\
&&\nonumber
{{n_\pm \beta \cos \left( \theta  \right)-\cos \left( {\phi_{\rm s}} \right)} \over
{\gamma \beta ^2\sin \left( {\phi_{\rm s}} \right)}}
{1 \over {g\left( {\phi_{\rm s}} \right)}}
{{{\rm d}g\left( \phi  \right)} \over {{\rm d}\phi }} \Bigr]
\end{eqnarray}

Where we corrected Ramaty's \shortcite{R} formula by deleting 
a factor $1+\partial \ln(n_{\rm ref}) / \partial \ln(\nu)$, and changed the minus 
sign multiplying the $J_{\rm s}^\prime$ term because 
our $T_\pm$ have the opposite sign to Ramaty`s $a_{\theta\pm}$.
We also added the longitudinal contribution which is important for 
frequencies close to the cyclotron frequency. 
$J_{\rm s}$ are Bessel functions, $J_{\rm s}^\prime$ their derivatives,
with the argument
$x_{\rm s}=(\nu/\nu_{\rm B}) n_\pm\gamma\beta\sin(\theta)\sin(\phi_{\rm s})$. 
The limits of the summation are
$s1,2={{\gamma \nu}\over{\nu_{\rm B}}} [1 \pm n_{\rm ref} \beta \cos(\theta)]$,
and $\pm$ designates the mode ('+' for OM and '-' for XO).
The angle $\phi_{\rm s}$ comes from the resonance condition (Ramaty 1969; Melrose 1989)
\begin{equation}
\label{resonance}
{{s\nu_{\rm B}}\over \gamma}=\nu [1-n_{\rm ref}\beta\cos(\theta)\cos(\phi)]
\end{equation}
and is derived by
\begin{equation}
\cos(\phi_{\rm s})={ {1- { {s\nu_{\rm B}}\over{\gamma \nu}}}\over{n_\pm \cos(\theta) \beta} }
\end{equation}

 The distribution function of the electrons was assumed to be separable into the form
\begin{equation}
\label{dist}
f({\bmath p}){\rm d}^3p=2\pi u(\gamma)g(\phi){\rm d}\gamma {\rm d}(\cos(\phi))
\end{equation}
Where $\phi$, the pitch angle, is the angle of the velocity to the magnetic field,
and $\gamma=(1-\beta^2)^{-0.5}$ the Lorentz factor. \par
  It is important to note that the equation for the absorption coefficient 
has the same properties as the equation for the growth rate since 
the growth rate is 
\begin{equation}
\label{growth}
\Gamma_\pm=-k_\pm\cdot v_{\rm g}
\end{equation}
 where $v_{\rm g}$ is the group velocity given by \cite{MelroseB}
\begin{equation}
\label{vg}
 {\bmath v}_{\rm g} = {{\partial\omega}\over{\partial k}}=
{c\over {{\partial(\omega n_{\rm ref})}\over{\partial\omega}}}
\left(\hat k-{{L_\sigma T_\sigma}\over{1+T_\sigma^2}} {\bmath t}\right)
\end{equation}
 As in all previous studies, we neglected the angle between the group velocity
and the wave vector. The group velocity in our study is therefore assumed
to be in direction $\hat k$, and of magnitude
\begin{equation}
\nonumber
v_{\rm g} ={c\over {{\partial(\omega n_{\rm ref})}\over{\partial\omega}}}
\end{equation}
\par
  In this work we assumed the power-law electrons have a power-law dependence
on kinetic energy 
\begin{equation}
\label{powerlaw}
u(\gamma)=n_{\rm hot}\cdot(\delta-1)\cdot(\gamma_{\rm min}-1)^{\delta-1}\cdot 
(\gamma-1)^{-\delta}
\end{equation}
where $n_{\rm hot}$ is the number density of the fast electrons, and
the power law starts  from a low-energy cut-off $\gamma_{\rm min}$
and ends at a high-energy cut-off $\gamma_{\rm max}$.
The distribution is normalized to $n_{\rm hot}$ when we neglect
$(\gamma_{\rm min}-1)/(\gamma_{\rm max}-1)$.
For the ambient plasma we used a thermal distribution 
\cite{Lang}
\begin{equation}
\label{thermal}
u(\gamma){\rm d}\gamma=n_{\rm cold} {2 \over \sqrt{\pi}}
\left({{mc^2}\over{kT}}\right)^{3/2} \sqrt{\gamma-1}\times
                        {\rm e}^{-{{mc^2}\over{kT}}(\gamma-1)} {\rm d}\gamma
\end{equation}
where $n_{\rm cold}$ is the number density of the cold, ambient plasma, electrons. \par
  For the pitch angle dependence in the fast distribution we used an ideal loss-cone
\cite{D+M}
\begin{equation}
\label{losscone}
g(\phi)=\cases{ Const., &$\cos(\phi) < \cos(\alpha)$ \cr
               Const.*{ {cos(\alpha-\Delta\alpha)-cos(\phi)} \over {cos(\alpha-\Delta\alpha) - \cos(\alpha)} }, &$\cos(\alpha) < \cos(\phi) < $ \cr
&$\cos(\alpha-\Delta\alpha)$ \cr
                        0, & $\cos(\alpha-\Delta\alpha) < \cos(\phi)$ \cr}
\end{equation}

  This describes an isotropic distribution for pitch angles above some $\alpha$,
no electrons below another angle $\alpha - \Delta \alpha$, and a linear in
$cos(\phi)$ decrease within the $\Delta \alpha$ interval. 
The normalizing constant in $g(\phi)$ depends on the loss cone parameters. 
For an idealized loss cone as above it is 
$Constant =1 / \pi /(2+\cos(\alpha)+\cos(\alpha-\Delta\alpha))$.
With this definition the integral is normalized so that
$2\pi \int_{-1}^{1} g(\phi) {\rm d}(\cos\phi)= 1$. \par
  For the thermal plasma the distribution is isotropic and therefore
$g(\phi)=1/4/\pi$. \par

\section{Other Processes}
   There are other processes in a thermal plasma which have to
be taken into account.
  The free-free absorption coefficient for $T > 3\cdot 10^5$ K 
is \cite{Lang}

\begin{eqnarray}
\label{freefree}
k_{\rm ff}= 0.01 {{n_e^2}\over {\nu^2 T^{3/2}}} \left(24.57+ \ln{{T}\over {\nu}} \right)
\end{eqnarray}
where $n_e$ is the density of the thermal electrons in ${\rm cm^{-3}}$, 
$\nu$ the frequency of the absorbed radiation in Hz, and $k_{\rm ff}$ the 
absorption coefficient in ${\rm cm^{-1}}$ . \par
  Our assumed number density is of order $n_e \le 10^{11}$,  our standard
temperatures are of order $T \approx 10^7$, and the emission frequency for
our standard values is of order $\nu \approx 1 {\rm GHz}$.
The free-free absorption for these values is $k_{\rm ff} \approx 5\cdot 10^{-8}$.
For a standard number of fast electrons of order  
$n_{\rm hot} \approx 10^6\ {\rm cm^{-3}}$ the free-free absorption normalized to
one electron is $k_{\rm ff}/n_{\rm hot} <  5\cdot 10^{-14}$,  which is negligible
compared to the gyrosynchrotron absorption
of the dominant modes. \par
   The deflection time for a particle in a thermal
plasma, with a particle velocity much higher than the thermal velocity but
still non-relativistic, is \cite{BenzBook}
\begin{equation}
t_{\rm d} = 3\cdot 10^{-20} {{v^3}\over{n_e}}
\end{equation}
where $v$ is the particle velocity, and $t_{\rm d}$ is the time for a deflection
of $90$ degrees. \par
  Taking an electron with $\gamma=1.02$, which is the low energy cutoff of
our fast electron distribution,
the deflection time is $t_{\rm d} \approx 62$ msec. The time for a deflection
of only a few degrees is much shorter, but the timescale of 
saturation is a few milliseconds, and therefore the deflection should
not be an important factor \cite{Aschwanden b}.
Faster particles have even longer 
deflection times, and are even less affected. \par

\section{Conditions for Maser and Saturation Length}

   Inspection of equation \ref{gyroe} shows that the absorption
for a distribution which is isotropic in pitch angle is negative only if
the slope of $u(\gamma)/\gamma^2/\beta$ is positive. This is the equivalent
of an increase of the fast particles with energy, a distribution which is
not likely to occur. The other possibility for amplification is when the distribution
is not isotropic in pitch angle. In that case if the derivative of $g(\phi)$ is large 
enough, and there are solutions such that the pitch angle dependent term 
in equation \ref{gyroe} is negative, then the absorption will be negative. 
One such distribution is the loss-cone distribution, which we use. 
   The loss-cone distribution arises naturally in a solar flare where the magnetic 
field and the density increase with decreasing height from the corona to the 
chromosphere, and we expect a loss cone to appear near the foot-points of a flare. \par
     A numerical computation of equation \ref{gyroe} with  a loss-cone 
(equation \ref{losscone}) and a power-law (equation \ref{powerlaw}) results
in negative absorption coefficients for several frequencies and angles to the magnetic 
field. Waves of these frequencies, and propagating with these angles,  
are amplified, and this is the Electron-Cyclotron-Maser mechanism. \par 
   The major factor determining which mode is the most important, or dominant, 
is $k\times L$, or $\Gamma\times L/v_{\rm g}$, where $L$ is the length of 
amplification \cite{BenzBook}. If $L$ is determined by 
homogeneity length-scales
in the loop then the mode with largest absolute value negative absorption is dominant.
However, if the maser stops operating after a distance smaller than the
homogeneity scales of the loop, the amplification length is intrinsic to
the ECM itself. Our goal was to derive the absorption coefficients, and the length of 
amplification $L$. \par
  The intrinsic length of the maser was estimated with a model for the evolution of 
the fast electron distribution. We assumed the electrons 
are accelerated for a short time at the loop top, so that a pulse of fast electrons 
with a duration of a few milliseconds to a few hundred milliseconds is produced. 
The fast electrons reach an abrupt increase in the magnetic field, and the electrons
with large pitch angles are reflected back. 
   We assumed that before reaching the mirror the fast electron 
distribution is isotropic in pitch angle, and that all the electrons below
a critical pitch angle pass the field increase. These assumptions result in
the ideal loss cone distribution described by equation \ref{losscone} with 
a loss-cone opening angle of $\alpha$. 
 Our evolution model therefore began with an ideal loss-cone distribution over 
some volume, and at time $t=0$ the maser is ``turned on``. We then assumed that 
the anisotropic loss-cone distribution relaxes into a new distribution via the maser,
and for simplicity we assumed the new distribution is also 
an ideal loss cone. The new distribution has a smaller opening angle $\alpha\prime$,
and a power law in energy of the same power index as the original
distribution.  If the dominant process which causes the relaxation is the maser, the 
difference in energy between the initial loss-cone distribution and the final 
distribution is the energy available to the maser. \par
  A useful measure of the amplification process is the {\it saturation
length}, $l_{\rm sat}$.  We define the saturation length as the distance a wave can be 
amplified before the loss-cone distribution is relaxed so that the loss cone
disappears, or the particular wave is no longer amplified. 
  Complete relaxation for a volume element occurs when the energy in
the element is reduced from the anisotropic distribution level to the
isotropic distribution level.  Our method for calculating the saturation length 
assumes that the greatest possible amplification 
is when the entire energy difference between initial and
final loss-cone configurations is turned into maser radiation. 
We estimate this energy difference by assuming that the relaxation process leaves
the number of fast electrons and the total parallel momentum unchanged.
The length of amplification can not be larger than the smallest saturation
length, and therefore if $l_{\rm sat}$ is smaller than the loop length-scales 
it determines the dominant mode.  The mode with the smallest saturation
length is therefore the dominant mode. \par
  We followed the gradual closing of the loss cone in  detail for some test cases, 
and our results justify an all-or-nothing approximation. In this approximation we assume
that the distribution remains unchanged until the energy radiated by the maser is 
equivalent to some per cent of the kinetic energy, and then the relaxation to the lower
energy configuration is instantaneous. When the closing of the loss-cone is followed
it is found that the loss-cone opening angle is usually changed only by a few degrees
before the absorption coefficient for the mode which had the largest 
amplification becomes positive. However, this happens after a distance which is only 
larger by $20$ to a $100$ per cent of the $l_{\rm sat}$ calculated with the all-or-nothing 
approximation, when we assume the energy turned into the maser is $10$ per cent of the 
kinetic energy.  The energy difference with the gradual closing method is usually 7--10 
per cent of the total kinetic energy.  Our all-or-nothing method is simple and gives a transparent 
relation between the absorption and the saturation length. \par
\begin{figure}
  \plotone{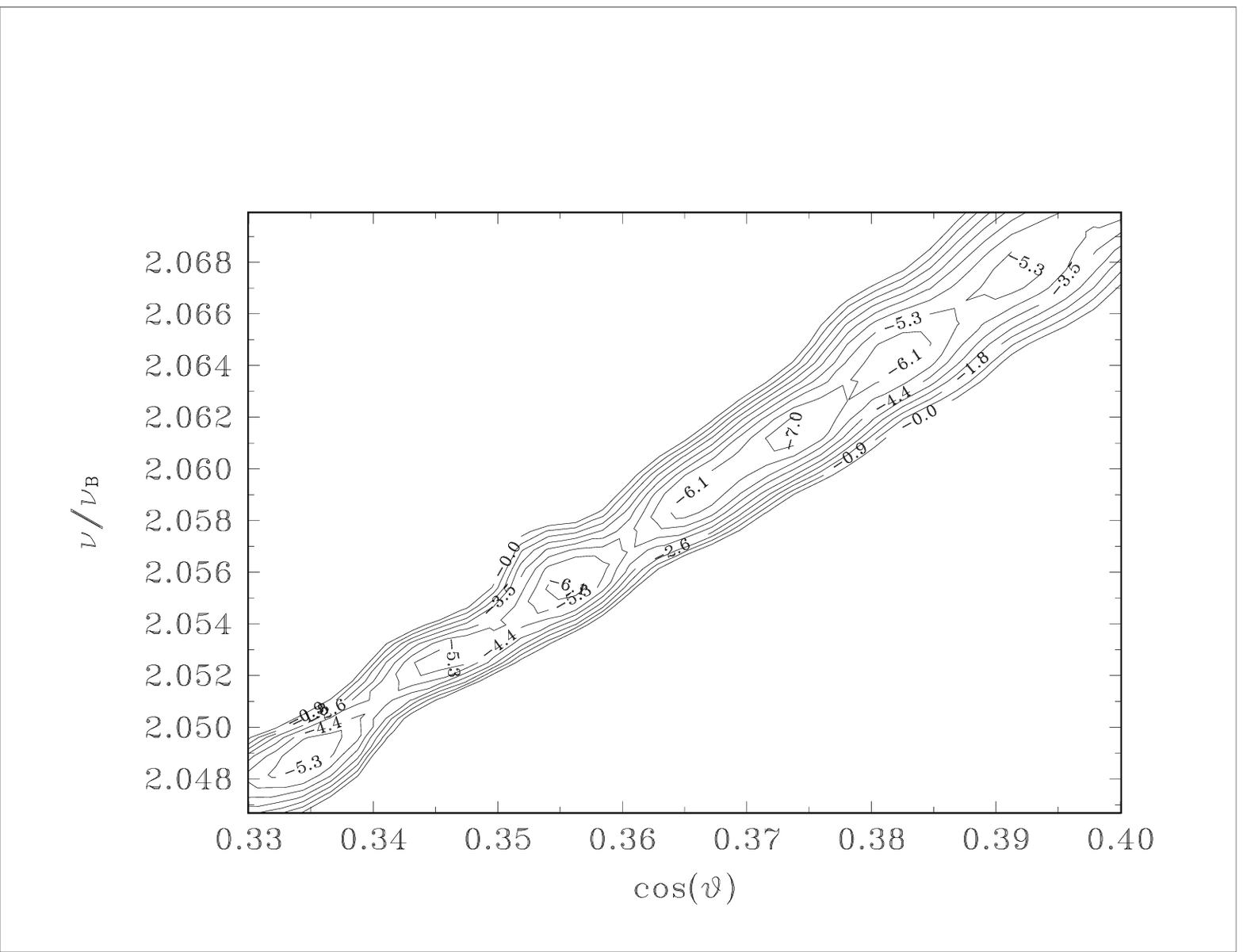}
\caption{Contour map  of the area surrounding the maximum
negative absorption in the XO mode for $\nu_{\rm p}/\nu_{\rm B}=1$,
temperature $T=5\cdot 10^6$ Kelvin, and $n_{\rm cold}/n_{\rm hot}=10^4$.
The x-axis of the figure denote $\cos(\theta)$, the y-axis denote the normalized 
frequency $\nu/\nu_{\rm B}$, and the absorption coefficients are normalized to $n_{\rm hot}=1$, 
and multiplied by $-10^{12}$}
\label{XOT=5}
\end{figure}
\begin{figure}
  \plotone{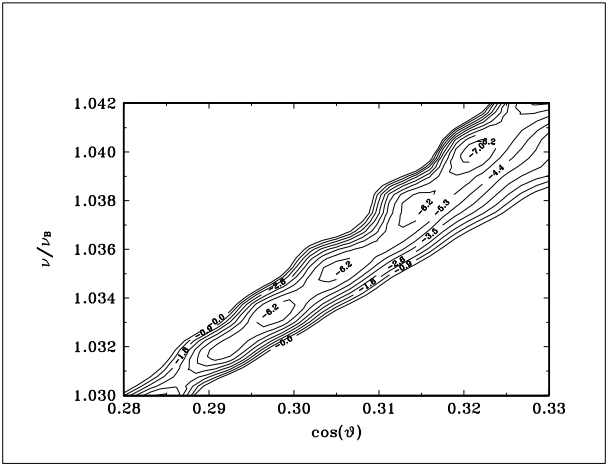}
\caption{Contour map  of the area surrounding the maximum
negative absorption in the ZM for $\nu_{\rm p}/\nu_{\rm B}=1$,
temperature $T=5\cdot 10^6$ Kelvin, and $n_{\rm cold}/n_{\rm hot}=10^4$.
The x-axis of the figure denote $\cos(\theta)$, the y-axis denote the normalized 
frequency $\nu/\nu_{\rm B}$, and the absorption coefficients are normalized to $n_{\rm hot}=1$, 
and multiplied by $-10^{12}$}
\label{ZMT=5}
\end{figure}

   Our calculations are predicated upon the assumptions that the magnetic field 
strength $B$, the direction to the observer $\theta$, and 
the densities  $n_{\rm cold}, n_{\rm hot}$ are constant. Figures
\ref{OMT=5}, \ref{XOT=5}, and
\ref{ZMT=5} show contours of negative absorption coefficients around the area of
maximum negative absorption. From these figures it is evident that a change in direction
of the magnetic field by even $1\degr$ removes a propagating wave from the 
peak of negative absorption. Likewise a change of 1 per cent in the magnetic field
makes the normalized frequency $\nu/\nu_{\rm B}$ of a wave change by about 1 per cent,
and also removes the wave from the peak of amplification.
Estimates of the scale lengths for the magnetic field are usually 
such that changes of $1$ per cent occur over distances of about $L=10^7$ cm
\cite{B+A2}.
Therefore, for saturation lengths much larger than $10^7$ cm, we 
expect the amplification ceases before the particular wave reaches its saturation
length, and the loss-cone either persists through the loop, or perhaps relaxes by some other means.

\subsection{Energy in the Loss-Cone}
\label{E in loss-cone}
  The total momentum parallel to the magnetic field for a one-sided loss cone 
distribution with constant density in angle up to $\cos(\phi)=\cos(\alpha)$ and no
electrons for higher $\cos(\phi)$, and with power law in kinetic energy is
\begin{eqnarray}
\label{P_par}
&&P_{||} = \\
&&\nonumber
 \int_{\gamma_{\rm min}}^{\gamma_{\rm max}}\int_{\phi=0}^{\phi=\pi/2}
{\rm d}\gamma {\rm d}(\cos(\phi))
m\gamma\beta c \cos(\phi) 2\pi u(\gamma)g(\phi)=\\
&&\nonumber
 = { {mc} \over {2\pi(1+\cos(\alpha))} } { {\cos^2(\alpha)} \over 2 } W_{\gamma}
\end{eqnarray}
  Where the dependence on energy is included in the integral
\begin{eqnarray}
W_{\gamma} = n_{\rm hot} (\delta - 1) (\gamma_{\rm min} -1 )^{\delta-1}
\int{\sqrt{\gamma^2-1} (\gamma - 1)^{-\delta} {\rm d}\gamma}
\end{eqnarray}
  and $\gamma_{\rm min}$ is the lowest energy, $\gamma_{\rm max}$ the highest 
energy, of the fast electrons. \par
  The dependence on the opening angle of the loss-cone $\alpha$ is
included in the first part of the formula for $P_{||}$, where the integral over $\phi$ 
was done on the positive side of the distribution,  $\phi \leq 90$, and
where we assume the
transition zone from isotropic distribution to the empty region is negligible. \par
   The energy dependent integral $W_{\gamma}$ is especially simple for the case of 
a power law in energy with $\delta=3$ and is
\begin{eqnarray}
\label{delta=3}
W_{\gamma}= {{2 n_{\rm hot}} \over 3} ( 1+ \gamma_{\rm min})^{3/2}
(\gamma_{\rm min}-1)^{1/2}
\end{eqnarray}
where we have neglected, as is usually done, terms of order
 $ (\gamma_{\rm min}-1) / (\gamma_{\rm max}-1)$. We use $\gamma_{\rm min}=1.02$ and $\gamma_{\rm max}\geq 3$ in our calculations. \par
We also need the equation for the kinetic energy density in a power law
distribution which is
\begin{equation}
\label{Ek}
E_{\rm k} = { {\delta -1} \over {\delta-2} } mc^2 \ n_{\rm hot} \ (\gamma_{\rm min} -1) \left[{\rm erg\ cm^{-3}}\right]
\end{equation}
where $n_{\rm hot}$ is the number density of fast electrons. \par
  We demand conservation of parallel momentum when the distribution
changes from a loss-cone with opening angle $\alpha$ and $\gamma_{\rm min}$
to a distribution with $\gamma_{\rm min} \prime$ and $\alpha \prime$, so that
$P_{||}(\cos(\alpha),\gamma_{\rm min}) = P_{||}(\cos(\alpha\prime),\gamma_{\rm min} \prime)$.
  From the parallel momentum conservation condition, and given
$\cos(\alpha),\gamma_{\rm min},\cos(\alpha\prime)$, we calculate $\gamma_{\rm min} \prime$. 
From $\gamma_{\rm min} \prime$ and equation \ref{Ek} we calculate 
the amount of energy lost in the transition from a loss cone with $\alpha$ to a loss
cone with opening angle $\alpha\prime$.
  The transition to a lower cut off energy is physically reasonable
since we expect electrons with low $\gamma$ to emit in the low
frequencies more than electrons with high $\gamma$, and when these
electrons lose energy they move to smaller $\gamma$-es. \par
  As an example we took an initial distribution with $\cos(\alpha)=0.8, 
\gamma_{\rm min}=1.02$, which is similar to our standard loss-cone.
From equations \ref{P_par} to \ref{delta=3} the isotropic distribution 
($\cos(\alpha\prime)=1$) which has the same 
total parallel momentum is a distribution with $\gamma_{\rm min} \prime = 1.01026$. 
   Therefore, about fifty per cent of the energy was
lost in the transition from a loss-cone with $\cos(\alpha)$=0.8 and 
$(\gamma_{\rm min}-1)=0.02$, to an isotropic distribution with $\cos(\alpha \prime)=1$ 
and $(\gamma_{\rm min} \prime-1)=0.01026$.
An estimate of fifty per cent is two orders of magnitude larger than that
found by Mackinnon, Vlahos \&Vilmer \shortcite{Mackinnon} who estimated 
the energy as 1 per cent of total kinetic energy, and one
order of magnitude larger than that given by
White, Melrose \& Dulk \shortcite{WMD}.  However, this estimate is about the same
as quoted by White et.al. \shortcite{WMD} as resulting from analytic
arguments. \par
  We also expect that a particular wave is no longer amplified
once the loss cone characteristics have changed even slightly. Therefore 
we should have calculated the energy loss when the loss cone angle changed
from $\cos(\alpha)=0.8$ to, say,  $\cos(\alpha\prime)=0.82$. The energy
loss for this case turns out to be only seven (7) per cent.
  Since our purpose was merely to get an estimate for the energy available to the 
maser, we can take as a reasonable estimate ten per cent of the kinetic energy. 
The exact number has very little effect on the saturation length, but the brightness
temperature is directly proportional to the energy (see equation \ref{TB2}). \par

\subsection{Saturation Length}
\label{saturation-length}
  To calculate the saturation length from the energy estimated to be available to 
the maser we assume that the maser is dominated by 
one mode $(\nu,\theta)$. We assume that the entire energy is 
concentrated in this one frequency and direction, with a typical angular width 
according to our numerical computations of  $\Delta\Omega = 0.1\ \pi$, and a 
typical frequency width of $\Delta\nu=0.015\nu_{\rm B}$. \par 
  To decide which is the most important mode we used 
the typical frequency and angular widths, calculated the saturation length for every 
mode as if that mode was the dominant one, and then compared saturation lengths. 
Usually the mode with the largest absolute value absorption coefficient also has the 
smallest saturation length, and is therefore the dominant mode. \par
\begin{figure}
  \plotone{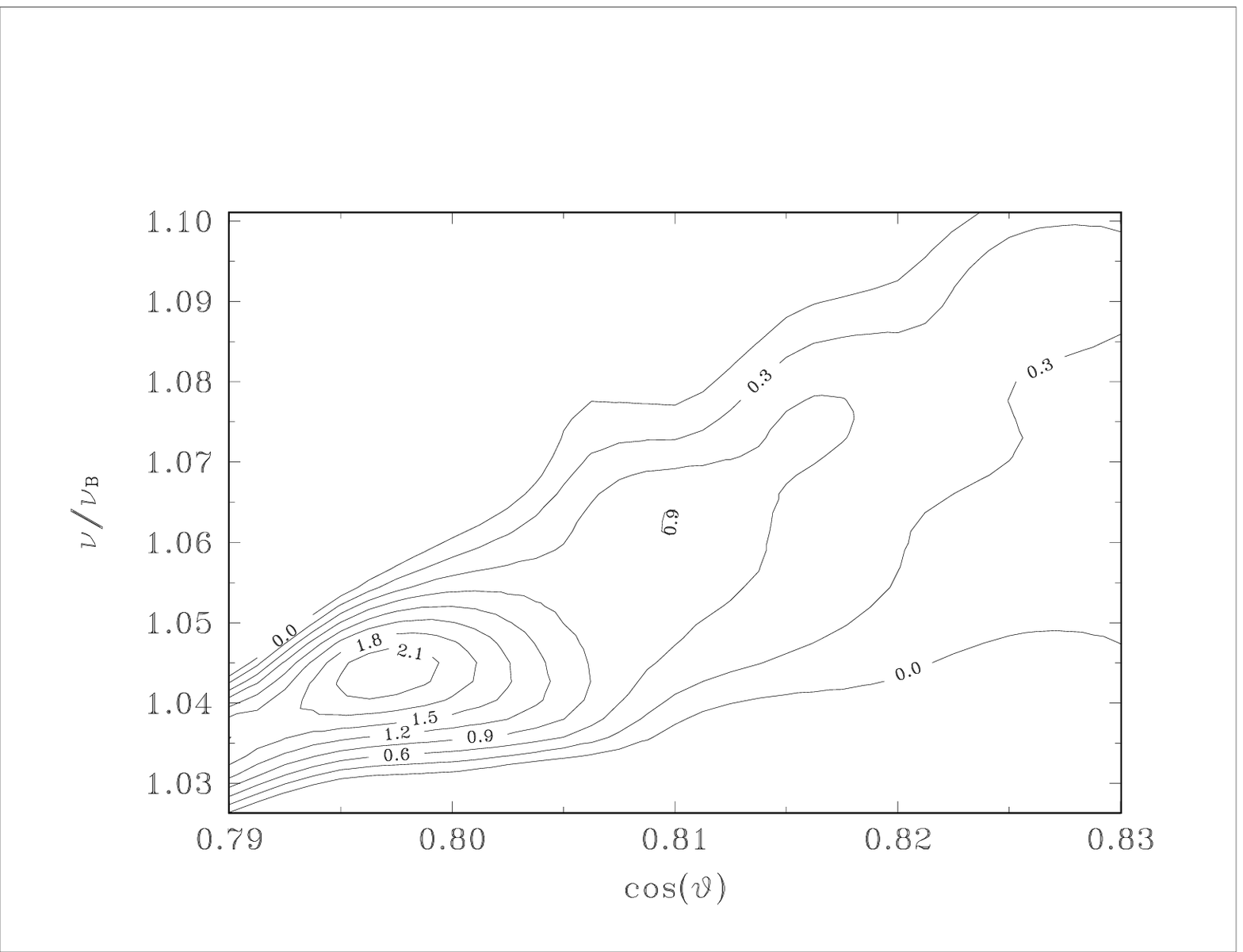}
\caption{Contour map  of the area surrounding the maximum
negative absorption in the OM for $\nu_{\rm p}/\nu_{\rm B}=1$,
temperature $T=5\cdot 10^6$ Kelvin, and $n_{\rm cold}/n_{\rm hot}=10^4$.
The x-axis of the figure denote $\cos(\theta)$, 
the y-axis denote the normalized 
frequency $\nu/\nu_{\rm B}$, and the absorption coefficients are normalized to $n_{\rm hot}=1$, 
and multiplied by $-10^{12}$}
\label{OMT=5}
\end{figure}

 A justification of this approach is seen for example in Fig.
\ref{OMT=5}.
The figure shows contours of equal absorption coefficient
for the case $\nu_{\rm p}/\nu_{\rm B}=1, T=5\cdot 10^6$, and the OM. The
absorption coefficients are negative, and therefore the figure shows equal
amplification contours.
The maximum amplification in Fig. \ref{OMT=5} is in the region
$0.795<\cos(\theta)<0.8$ and $1.04<\nu/\nu_{\rm B}<1.045$.  If the magnetic field 
strength decreases by one per cent the frequency of the wave moves from
approximately $1.04 \nu_{\rm B}$ to approximately $1.05 \nu_{\rm B}$, and the
amplification is decreased by approximately half.
Likewise, if the inclination of the field changes by one 
degree, the amplification is also reduced by approximately half. 
It is therefore justified to use narrow frequency and angular widths. \par
 The simplest method of calculating the saturation length is the all-or-nothing method.
The amplification of any mode is calculated from the length over
which there is negative absorption, which is assumed to be arbitrary at first. 
On the other hand the intensity
produced from a given volume must be limited by the total energy in the
volume. The length for which the loss cone persists 
is calculated self-consistently by equating
the energy in the wave, and the energy in the volume. \par
The total power of the maser per unit area is 
\begin{equation}
P  = I(\nu,\cos(\theta)) \Delta\Omega \Delta\nu
\end{equation}
 
 The energy increases over a volume equal to the path of amplification 
times unit area, and therefore the total energy available to the maser is
\begin{equation}
0.1\times E_{\rm k}\times l_{\rm sat}\ \ [{\rm erg\ cm^{-2}}]
\end{equation}
   where we used section \ref{E in loss-cone} and our estimate of 10 per cent of the 
total kinetic energy as the energy available to the maser, and $l_{\rm sat}$  is the 
length of amplification -- which is unknown as yet. \par
   Assuming that the maser lasts for a given time $\Delta t$, and that 
the intensity of the emission is constant for that time, the equation 
which has to be solved is
\begin{eqnarray}
P \Delta t =0.1\times E_{\rm k}\times l_{\rm sat}
\end{eqnarray}

 By using the radiation transfer equation for negative absorption
\begin{equation}
\label{radtran}
I = {j\over k} (1-{\rm e}^{-k x}) \approx {j\over {|k|}} {\rm e}^{|k|x}
\end{equation}
where we have neglected $1$ compared to the exponent, and assumed that
$j,k$ are constant everywhere, we have 
\begin{eqnarray}
\label{lsat2}
&&{j\over{|k|}} {\rm e}^{n_{\rm hot} |k| l_{\rm sat}} \Delta\Omega \Delta\nu \Delta t = \\
&&\nonumber
0.1\times 2\times 8.2\cdot 10^{-7}\times n_{\rm hot}\times (\gamma_{\rm min} -1) 
\times l_{\rm sat} 
\end{eqnarray}
   where we used equation \ref{Ek} with $\delta=3$, and also introduced 
normalized (to $1$ particle) absorption and 
emission coefficients. \par 
 If we assume the maser lasts for a given time $\Delta t$ the 
equation can be solved numerically for any $\Delta t$. 
  However, if we assume $\Delta t=l_{\rm sat}/v_{\rm g}$, with $v_{\rm g}$ given 
by equation \ref{vg}, 
then the solution is straightforward, and the result of the simple method is 
\begin{equation}
\label{satlen}
l_{\rm sat} = \ln\left({{0.1\cdot E_{\rm k}\cdot v_{\rm g}\cdot |k|}\over
{\Delta\Omega\Delta\nu\cdot j}}\right) /(n_{\rm hot} |k|)
\end{equation}

   A more complicated method of deriving the saturation length is the
differential approach.
We assume a loss-cone distribution over some volume of space.
We further assume that the maser is ``turned on`` at time $t=0$.
The intensity at time $t$ arriving into a volume element
is given by equation \ref{radtran} with $x=v_{\rm g} t$, since a wave starting at 
time $t=0$ has advanced this distance by time $t$.
   A volume element adds to the intensity a fractional 
intensity 
\begin{equation}
\nonumber
{\rm d}I=(j-kI){\rm d}x
\end{equation}
and again neglecting $1$ compared to the exponent, the intensity
increase for a volume element is
\begin{equation}
{\rm {dI}\over{dx}}\approx |k|I\approx j {\rm e}^{|k|v_{\rm g}t}
\end{equation} 
The power in ${\rm erg\ s^{-1}}$ drawn from a unit volume at time $t$ after
the maser was turned on, assuming that $k$ and $j$ remain independent of $t$ and the same everywhere,
is therefore:
\begin{equation}
\Delta P(t) \approx \int {\rm d}\Omega \int {\rm d}\nu
j {\rm e}^{|k|v_{\rm g} t}
\end{equation}
The energy drawn from from a unit volume from time
$t_1$ to time $t_2$ is
\begin{equation}
\Delta E=\int_{t_1}^{t_2}\Delta P(t) {\rm d}t \approx
\int {\rm d}\Omega \int {\rm d}\nu
{{j}\over{|k|v_{\rm g}}}({\rm e}^{|k|v_{\rm g} t_2}-{\rm e}^{|k|v_{\rm g} t_1})
\end{equation}

Since energy is being drawn from the volume element the 
distribution function changes, and here we use section \ref{E in loss-cone}.
We assume that the distribution changes so that it
remains a power-law with index $\delta$ in energy, and 
a loss-cone.
The new distribution function, with less
energy, has a new and smaller $\gamma_{\rm min}\prime$ and a 
new and larger $\cos(\alpha\prime)$.
From our calculated energy loss for some interval $t_2-t_1$
we write $E_{\rm k}^\prime=E_{\rm k} - \Delta E$, and derive $\gamma^\prime$.
From the condition that the parallel momentum is constant
$P\prime_{||}=P_{||}$, and using $\gamma^\prime$, we derive $\alpha^\prime$.
We can then calculate the new absorption and emission coefficient,
the new energy loss, and so on, and repeat the process iteratively. 
 If we take $t_1=0$ and $t_2=\tau=l_{\rm sat}/v_{\rm g}$, 
and substitute for the integrals the constants $\Delta\Omega$ and $\Delta\nu$, 
the equation for the saturation length we derive from the differential 
approach is identical to the equation of our simple method (equation \ref{satlen}). \par
    We followed the gradual closing of the loss-cone iteratively for a few test cases. 
We took into account the change of the absorption and
emission coefficients for waves arriving in later times, which passed through 
regions where the distribution has already changed. 
  The results of this method are that about 3--7 per cent of the kinetic
energy is turned into maser radiation, mostly in the dominant mode,
that this happens after a time  $<2\cdot \tau$,
and that the intensity of the maser is very near to the intensity 
after amplification with the original maximum absorption 
over length $l_{\rm sat}=\tau v_{\rm g}$.
Thus the natural growth time of the maser
to maximum is $\leq 2\ \tau$ and the length over which
maser amplification is possible is $\leq 2v_{\rm g}\tau$,
and our simple method is a very good approximation. \par
   Since the results of following the closing of the loss-cone and just
assuming a fixed energy available to the maser are so close,
we used for the rest of our calculations a fixed 10 per cent energy availability.
The results given in this work were calculated with this assumption.

\section{Calculations}
  Our calculations were carried out within the frame of the
cold-plasma and the magneto-ionic approximations 
\cite{MelroseB}.
We therefore disregard any negative absorption for frequencies which 
do not obey either of the two cold-plasma criteria \ref{criteria1} 
and \ref{criteria2}, where ``much larger`` in these criteria is taken to be at 
least $10$ times larger. \par
 We calculated the absorption coefficient with equation \ref{gyroe}, and
the emission coefficient with the equation appropriate for it \cite{R}. 
We performed the calculation using a power-law distribution in 
equation \ref{gyroe} to get the absorption due to the fast electrons, 
and also using a thermal distribution in the equation to get the absorption
due to the ambient plasma, and the absorptions were then added to derive the total 
absorption.  The total emissivity was derived in the same manner.
The calculations were carried out without any approximations to the Bessel 
functions, and the full sum over the $s$ values was performed.
The integral over the pitch angle $\phi$ was performed using the fully
relativistic resonance condition \ref{resonance}, and the distribution
functions $f({\bmath p})$ were normalized to the number density of the fast
electrons  for the power law, and the ambient plasma  for the thermal
distribution. 
The parameters used for the calculation were
\begin{itemize}
\item $\delta=3$ ;
\item $\gamma_{\rm min}=1.02$ ;
\item $\gamma_{\rm max}=3$;
\item $\cos(\alpha)=0.81$ ;
\item $\cos(\alpha-\Delta\alpha)=0.83$ ;
\item $B=360$ gauss ;
\item $n_{\rm cold}/n_{\rm hot}=10^4$, unless otherwise indicated ;
\item The ambient density is taken to be
$n_{\rm cold}=1.24\cdot 10^{10} (\nu_{\rm p}/\nu_{\rm B})^2\ {\rm cm^{-3}}$, which 
gives $\nu_{\rm p}=1$ GHz for $\nu_{\rm p}/\nu_{\rm B}=1$.
\end{itemize}
We used $\gamma_{\rm max}=3$ because several tests we performed showed 
that electrons with higher kinetic energy do not contribute to emission
in the frequencies of  the maser. \par
  For every given pair of $(\nu_{\rm p}/\nu_{\rm B},T)$ we calculated the absorption and 
emission coefficients on a grid of $\cos(\theta)=0.02-0.92$, with a spacing
of $\Delta\cos(\theta)=0.02$. For every $\cos(\theta)$ 
we calculated the coefficients for the frequency range from $\nu=\nu_{\rm B}$ to 
$\nu=3\nu_{\rm B}$, or higher if required, and with a changing resolution. 
The initial scanning resolution was changed from $\Delta\nu/\nu_{\rm B}=10^{-4}$ for 
the small $\cos(\theta)$ to $\Delta\nu/\nu_{\rm B}=0.01$ for the larger $\cos(\theta)$. 
When negative absorption coefficients were found the resolution was
increased ten fold, and the calculation for the range where the absorption is negative
was performed with the higher resolution. In this way we are confident that no
negative absorption in the $(\cos(\theta),\nu)$ range was missed.
Sometimes calculation was performed with better resolution
in $\Delta\cos(\theta)$ or $\Delta\nu/\nu_{\rm B}$ if the absorption coefficients
changed too much with the coarser resolutions. We performed calculations for 
the entire range $\nu_{\rm p}/\nu_{\rm B}=0.1-3$, but present only examples. \par
  Negative absorption appeared for the OM and XO near the harmonics of the cyclotron 
frequency, and for the ZM also near the upper hybrid frequency. Therefore it
was usually not necessary to search the whole range $\nu_{\rm B}-2\nu_{\rm B}$
and the whole range $2\nu_{\rm B}-3\nu_{\rm B}$ to find the negative coefficients. \par
    Calculation of the saturation length was performed as described
in the section {\it Saturation Length}, with the assumption of constant
width in angle and frequency. The widths enter the saturation
length only logarithmically, so the result is only weakly dependent
on them. Typical numbers are $\Delta\Omega=0.1\pi$
and $\Delta\nu=0.015\nu_{\rm B}$. \par
   The tables summarizing our results are arranged according to $\nu_{\rm p}/\nu_{\rm B}$.
Every row gives the characteristics of the most important wave for the given mode
and temperature, and for the $\nu_{\rm p}/\nu_{\rm B}$ of the table. The waves are
either those with the largest (in absolute value) absorption coefficient,
or those with largest growth rate, for all $\theta$ and $\nu$.
  The first column is the temperature used to calculate the absorption 
coefficient for the ambient plasma, in units of $10^6$ K. The
temperature affects only the thermal absorption and not the absorption
calculated from the power law electrons.
The second column is the frequency in units of the cyclotron 
frequency $\nu_{\rm B}$.
The third column is the emission angle to the magnetic field, or
the angle of the wave vector ${\bmath k}$ (assumed to be in the
$(x,z)$ plane) to the z-axis ( $\bmath B$ was assumed to be in the $z$ direction).
The fourth column gives the absorption coefficient itself, normalized
to a number density of the fast particles of $1$, and given
for a magnetic field of $360$ gauss which gives $\nu_{\rm B}=1$ GHz.
The absorption coefficient is therefore $n_{\rm hot}$ times the  $k$ which is 
given in the table.
The fifth column is the saturation length in units of $10^6$ cm,
computed with our formula \ref{satlen}, and the assumption that
$n_{\rm hot}=1.24\cdot 10^6\ (\nu_{\rm p}/\nu_{\rm B})^2\ {\rm cm^{-3}}$.
The sixth column is the group velocity, the seventh is the growth
rate, and the last column gives the mode for which these parameters
were computed. \par
  The contour figures show only negative absorption contours, multiplied
by a factor of $-10^{12}$ to make the numbers easier to display.
These are examples showing only the region of $(\cos(\theta),\nu/\nu_{\rm B})$ plane
around the largest in absolute magnitude negative absorption coefficient for
some mode.

\section{Observability of the Maser}
A previous study which used a power law in energy for the fast electrons found that the ZM
dominates for all $\nu_{\rm p}/\nu_{\rm B} > 0.5$ \cite{Fleishman2}, and
therefore concluded that the emission must come from regions with
smaller $\nu_{\rm p}$. Another study also concluded that the maser must
come from a region of low density $\nu_{\rm p}/\nu_{\rm B} < 0.25$, because for higher
densities the dominant modes reach saturation much faster than the observed
durations of MW spikes \cite{Aschwanden b}. \par
\begin{figure}
  \plotone{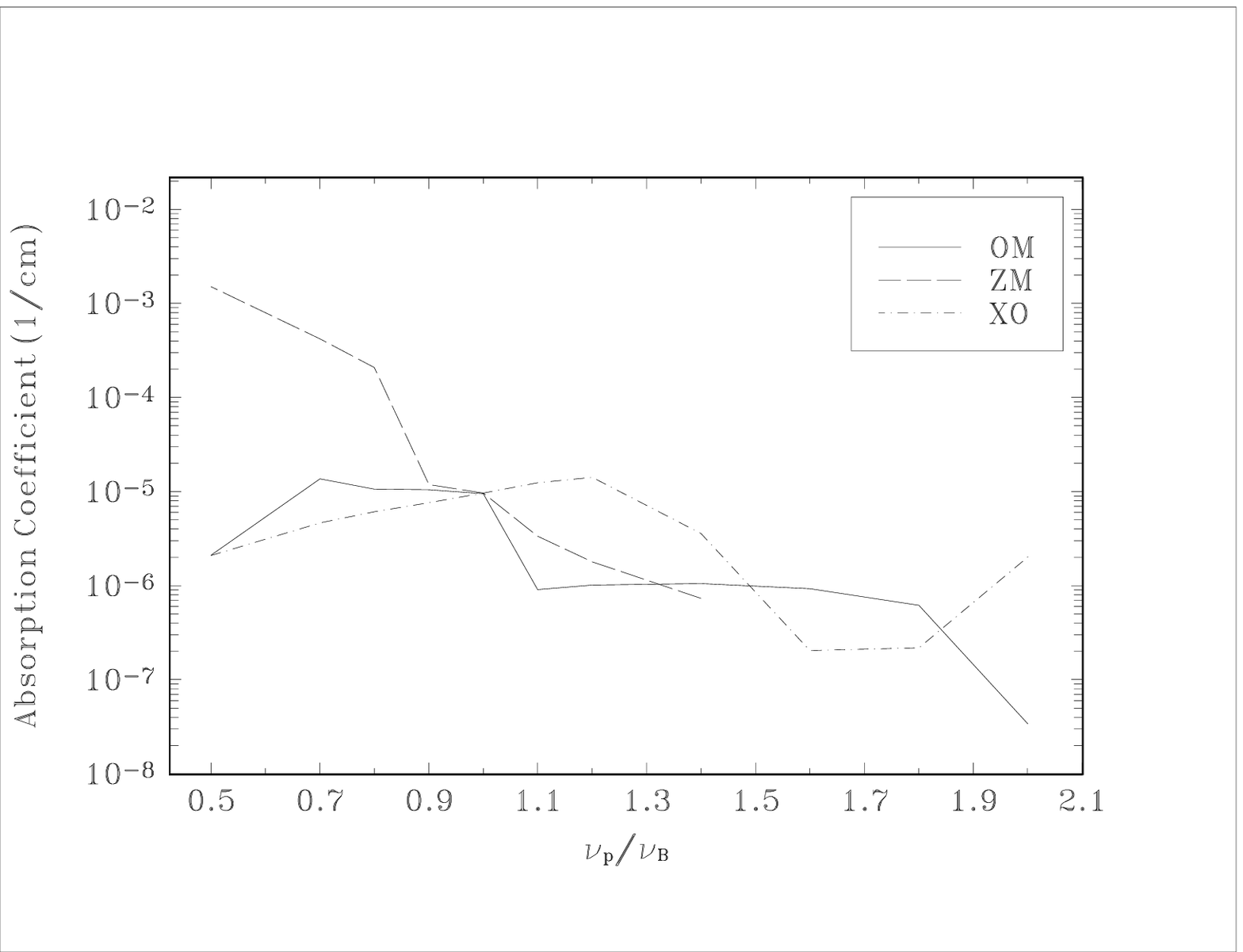}
\caption{Largest in absolute value negative absorption
  coefficient as function of 
$\nu_{\rm p}/\nu_{\rm B}$, for the ZM, OM and XO mode. The calculation was performed
with temperature of $5\cdot 10^6$ Kelvin, $n_{\rm cold}/n_{\rm hot}=10^4$,
power-law index of $\delta=3$, low-energy cutoff of $10$ keV, high energy
cutoff of $1$ MeV, loss cone parameters 
$\cos(\alpha)=0.81,\cos(\alpha-\Delta\alpha)=0.83$. 
The absorption coefficient given for any $\nu_{\rm p}/\nu_{\rm B}$ is the largest in the entire
frequency range $\nu=\nu_{\rm B}-3\nu_{\rm B}$, and angle to magnetic field range
$\theta=0-90$ degrees.}
\label{absorption}
\end{figure}
\begin{figure}
  \plotone{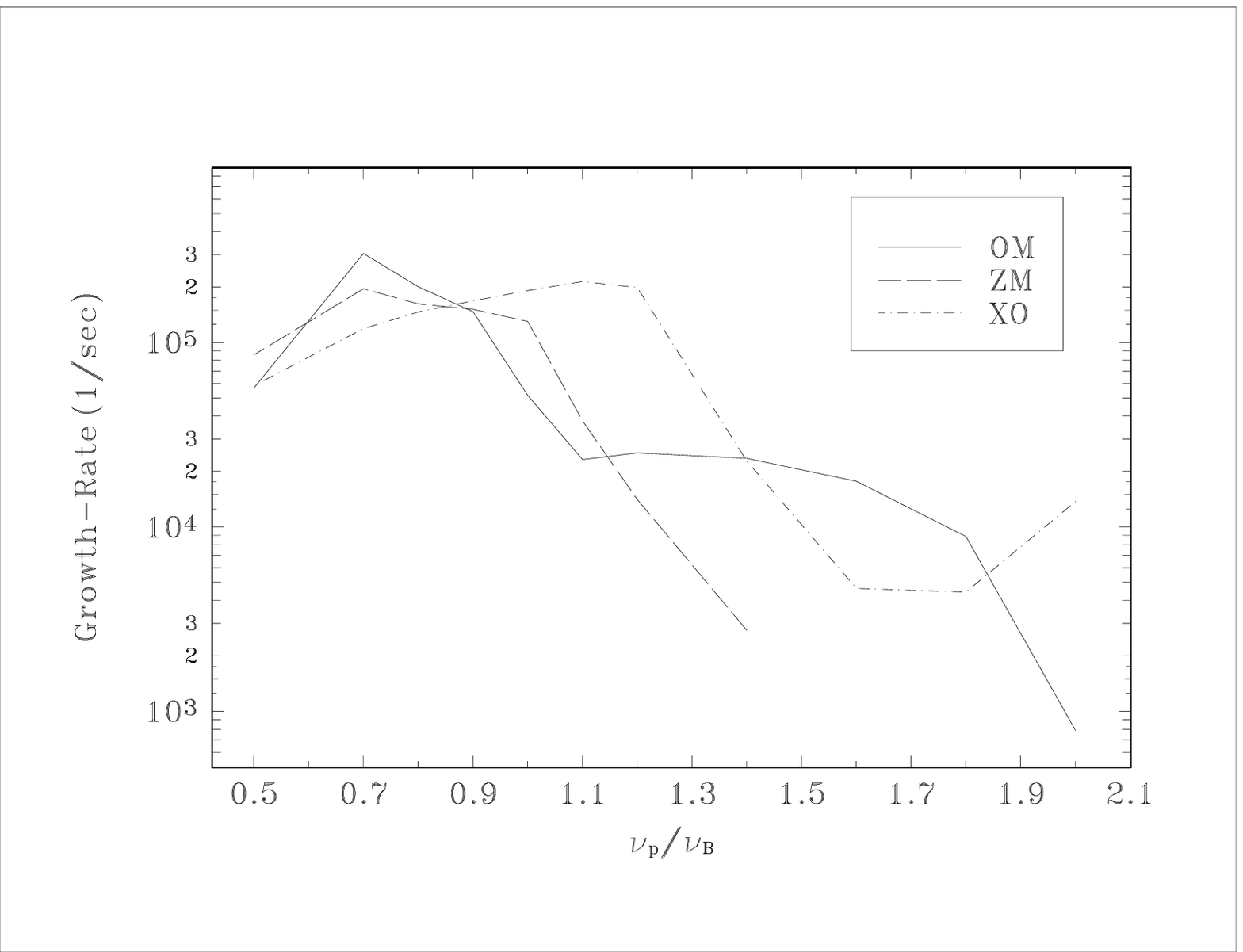}
\caption{Largest growth rate as function of 
$\nu_{\rm p}/\nu_{\rm B}$, for the ZM, OM and XO mode. The calculation was performed
with temperature of $5\cdot 10^6$ Kelvin, $n_{\rm cold}/n_{\rm hot}=10^4$,
power-law index of $\delta=3$, low-energy cutoff of $10$ keV, high energy
cutoff of $1$ MeV, loss cone parameters 
$\cos(\alpha)=0.81,\cos(\alpha-\Delta\alpha)=0.83$. 
The growth rate given for any $\nu_{\rm p}/\nu_{\rm B}$ is the largest in the entire
frequency range $\nu=\nu_{\rm B}-3\nu_{\rm B}$, and angle to magnetic field range
$\theta=0-90$ degrees. The largest growth rate does not necessarily occur
at the same angle and frequency as the largest absolute value negative
absorption.}
\label{growth-rate}
\end{figure}

     We found that for $\nu_{\rm p}/\nu_{\rm B}< 1.1$, and $\nu_{\rm p}/\nu_{\rm B} > 1.8$ the ZM has the 
largest in absolute value negative absorption (see Fig. \ref{absorption}), 
but has the largest growth rate only for $\nu_{\rm p}/\nu_{\rm B}< 0.6$ 
(Fig. \ref{growth-rate}).
   If the pulse of fast electrons has a much longer duration than the
time scale of maser saturation the important factor is not the growth
rate, but the saturation length.
At a time after the maser saturates, the mode with the smallest saturation length 
has closed the loss-cone beyond its own saturation length, and therefore the 
smallest saturation length becomes the amplification length for all modes. 
Modes with longer saturation lengths can reach only a fraction of the full 
amplification possible to them, being amplified by a factor $exp(|k|\cdot L)$, 
with $L$ the saturation length of the mode with the smallest 
saturation length, and not ${\rm exp}(|k|\cdot l_{\rm sat})$, with $l_{\rm sat}$ the saturation 
length of the mode which is $L<l_{\rm sat}$.
  Therefore what is required is to derive the saturation length for all modes,
and find the mode with the smallest saturation length. Since the saturation 
length is inversely proportional to the absorption coefficient, the mode with 
smallest saturation length is usually the mode with the largest 
(in absolute value) negative absorption coefficient. \par
  These findings present a problem in that we expect a loss-cone 
distribution to appear near the foot-points of flare loops where the magnetic 
field and the density increase sharply. 
The cyclotron frequency-plasma frequency ratio for the footpoints
is probably in the range $\nu_{\rm p}/\nu_{\rm B} \approx 0.5-3$, if we take as 
typical $n_{\rm cold} \approx 10^{10}-10^{11}\ {\rm cm^{-3}}$ for the density of 
ambient plasma \cite{B+A}, and 
$B \approx $ a few times $100$ gauss as typical for the magnetic field. 
These values are in
the range where our calculations, and previous work, predict
the unobservable ZM is the dominant mode.
   The high amplification of the ZM  would therefore seem to preclude the 
ECM mechanism as the producer of observable MW spikes, except in the 
range $1.1 < \nu_{\rm p}/\nu_{\rm B} < 1.8$. \par
   We have, however, to consider the ambient thermal plasma at the foot 
points, which we assume to be at a temperature of order $10^7$ K. 
The thermal effect was not taken into account in most studies, 
and those which did -- assumed much lower temperatures for the
ambient plasma than we do.  We use the range of temperatures inferred from SXR
observations, since we assume the model of a hot dense plasma, heated
by the fast electrons, flowing upwards to fill the loop, and emission
where the fast electrons intersect the evaporation front  \cite{B+A}. \par
  The temperature has two effects on our results. The first is to increase
the positive absorption near the harmonics of the cyclotron frequency.
If the frequency where the fast electrons create negative absorption 
is close to a harmonic, the thermal plasma absorption decreases
the amplification,
or even turns the negative coefficient positive. 
For the XO and OM this may eliminate the amplification at a low harmonic, 
but these modes
appear in higher harmonics and usually have negative absorption for the higher
harmonics as well. The ZM is limited to frequencies less than the upper-hybrid
frequency, and therefore it is possible that the introduction
of high temperatures will eliminate the ZM negative absorption while
leaving higher frequency OM and XO negative absorptions. 
  The second, and more important, consideration high temperatures 
introduce is the cold plasma conditions. For a temperature of $10^6$ K 
these criteria are valid for frequencies very close to the harmonics, 
and for very large refraction indices. 
However, the high temperature we assumed forced us to 
reject certain negative absorption results as not valid within the framework 
of our assumptions. We assumed that if there is any amplification for these cases
it is orders of magnitude smaller than the cold plasma result, and therefore
negligible. This is supported by the known effect of thermal corrections, which
is to decrease the refraction indice and the absorption \cite{MelroseB}. \par
   Even if the thermal positive contribution leaves a negative ZM 
absorption which is valid for the cold plasma approach,  and is the largest in 
absolute value of all negative absorptions,  the OM and XO may be observable.
   We assumed in the previous section that the largest in absolute value
negative absorption coefficient dominates, and that all the energy 
goes into amplifying this mode. With this assumption we derived the saturation
length and the amplification for this dominant mode.
This is similar to the assumption in previous studies, which give
the dominant mode in some parameter region as the mode with the
highest growth rate (White et.al. 1986; 
Aschwanden 1990a), and it is justified because 
of the exponential nature of the intensity.
  There are, however, two scenarios where modes with smaller absorption
coefficients can be observed. 
   The first scenario, presented in section \ref{non-dominant}, depends on 
the high brightness temperature 
of the maser. 
  The second scenario , presented in section \ref{slow-mode}, depends on the 
slowness of the Z-mode. 
If the ZM is slow enough, another mode may be able to reach its full 
amplification before the ZM saturates, and then, for a short time until
the ZM saturates, this mode can also be observed. \par
   Finally, there is a region where the ZM does not have negative absorption.
For the region $1.5\leq \nu_{\rm p}/\nu_{\rm B} \leq 1.6$  the lowest frequency for which the ZM 
exists is higher than $\nu_{\rm B}$ and the highest frequency for which the ZM 
exists is lower than $2\nu_{\rm B}$. Negative absorption usually appears near
the harmonics of the cyclotron frequency, and for the ZM frequencies
with this region of $\nu_{\rm p}/\nu_{\rm B}$ there are no negative absorption 
coefficients. \par
   All the above considerations are summarized in Fig. \ref{absorption} 
and in Fig. \ref{growth-rate} giving
the maximum absorption coefficients and growth rates for the modes.
The coefficients are the largest in all frequencies and angles for the
given $\nu_{\rm p}/\nu_{\rm B}$, wave mode, and with $T=5\cdot 10^6$ K. The ZM
does not appear for $1.6<\nu_{\rm p}/\nu_{\rm B}$ because where negative absorption
reappears at $\nu_{\rm p}/\nu_{\rm B} > 1.6$ the frequencies fail the cold plasma
criteria. 
It is important to note that the largest growth rate does not necessarily occur 
at the same angle and
frequency as the largest in absolute value negative absorption coefficient,
as can be seen in Tables \ref{ffp=0.8} to \ref{ffp=1.8}. \par
  Whatever mode is amplified, the ECM is difficult to observe because of 
the narrow bandwidth, and
especially because of the small angular width of the emission. 
Fig. \ref{OMT=5} shows contours of negative absorption coefficients around the 
area where the OM has the largest negative absorption (largest in absolute value). 
To observe the emission it is necessary to be within a cone of less than $1$ degree around
the angle of maximum amplification, and within $1$ per cent of the frequency of maximum
amplification. The absorption $1$ per cent away from the maximum drops by 
half, which for an exponential growth with a maximum power of the exponent
of about $30$ means 
a drop of $7$ orders of magnitude in amplification.
In Fig. \ref{XOT=5} and Fig. \ref{ZMT=5} the same narrow bandwidth
and small angular width are seen for the other modes, and the phenomena
occurs for all other $\nu_{\rm p}/\nu_{\rm B}$ values.
This general result means that observation
of ECM should be rare, even if ECM is very common and occurs at all times. \par

\subsection{Thermal Effect}
   In most cases the thermal plasma tends to suppress maser
emission, as was noted in previous studies
(Aschwanden 1990b; Kuncic \& Robinson 1993). 
  For example for $\nu_{\rm p}/\nu_{\rm B}=0.8$ (Table \ref{ffp=0.8}) 
and $T=5\cdot 10^6$ K,
the most important OM was found at frequency $\nu/\nu_{\rm B}=1.03$. At a higher temperature 
of $10^7$ K the most important OM moved to a higher frequency of $\nu/\nu_{\rm B}=1.079$.
The fast electrons still gave the same negative absorption at $\nu/\nu_{\rm B}=1.03$,
only for $T=10^7$ the thermal absorption was large enough to turn the absorption
at this frequency positive. On the other hand, the ZM at $\nu/\nu_{\rm B}=1.2745$ was not affected
by the thermal absorption, because of its distance from the cyclotron
frequency. The change there was the result of the application of the cold 
plasma criteria, which for $T=10^7$ this frequency violates, and therefore the 
largest acceptable absorption was at a lower frequency, further from the resonance 
and with a smaller refraction index.
  All these changes were not enough to change the dominant mode, which remained
the ZM, because the OM absorption was reduced by a factor of $8$, while
the ZM absorption was reduced only by a factor of $4$. 
The total effect of the 
higher temperature was therefore to reduce the relative importance of the OM 
emission. \par
    The other ZM given in Table \ref{ffp=0.8} is the ZM wave with the largest
growth rate. This is a wave at a much lower frequency, and a slightly
different angle than the wave with largest (negative) allowable absorption 
coefficient. This mode also changes with the increase in temperature
from  frequency $\nu/\nu_{\rm B}=1.0351$ for $T=5\cdot 10^6$ to frequency 
$\nu/\nu_{\rm B}=1.087$ for $T=10^7$. In this case the absorption was only
reduced by a factor of $2$, and the growth rate by about the same. This is an example
where the mode with the largest growth rate is not the dominant mode. \par
   A counter example is demonstrated in Table \ref{ffp=1.8}. 
The first row of Table \ref{ffp=1.8} gives a negative absorption coefficient for 
the frequency $\nu/\nu_{\rm B}=2.04885$, which is very close to the ZM resonance
for this case.
The refraction index for this frequency and for the angle $\cos(\theta)=0.06$ 
is $n_{\rm ref}=4.42$, which was large enough for this solution to barely violate 
the 1st plasma criterion (eq. \ref{criteria1}), giving a value only $9.5$ times 
larger than $0.001683/2$. This is not allowed, which is why the row is marked
as ZMN in the table.
Therefore, for $\nu_{\rm p}/\nu_{\rm B}=1.8$ the temperature leads us to discard the
ZM waves as unimportant, and leaves us with the OM as the dominant mode. \par
  When we made the calculations in Table \ref{ffp=1.8} with a temperature of
$T=10^7$ K, all the ZM negative coefficients disappeared entirely, overcome by
the thermal absorption, and therefore there was no need to even consider the 
cold plasma criteria.  \par
   It is important to note a problem with Table \ref{ffp=1.8} which is 
that the negative
absorption coefficients are very small. The saturation 
lengths were therefore very large, and may violate the assumption that the maser
is very small compared to the dimensions of the loop. For example for
$T=10^7$ K the saturation length for the OM was found to be
$l_{\rm sat}=1.8\cdot 10^8$ cm,
which is large enough for our assumption that the magnetic
field, the density, and the temperature are constant to be incorrect.
In such a case the maser does not reach its saturation intensity, and
the loss cone may relax by some other means. \par

\subsection{Non-Dominant Mode}
\label{non-dominant}
  Since the intensity $I$ is difficult to measure we would like to estimate 
the brightness temperature
\begin{equation}
\label{TB}
T_{\rm B}={{c^2}\over{2\nu^2}} {I\over{k_{\rm B}}} = 3.26\cdot 10^{18} 
{I\over{(\nu / {\rm GHz})^2}}
\end{equation}
We can express the intensity in terms of the saturation length, use equation 
\ref{satlen}, and write the intensity (equation~\ref{radtran}) as 
\begin{equation}
\nonumber
I={j\over{|k|}}\left({{0.1 E_{\rm k} v_{\rm g} |k|}\over{\Delta\Omega\Delta\nu\ j}}\right)
\end{equation}

  Using this, the brightness temperature for the dominant mode can now be 
written as
\begin{equation}
\label{TB2}
T_{\rm B}={{3.26\cdot 10^{18}}\over{(\nu / {\rm GHz})^2}}
     {{0.1 E_{\rm k} v_{\rm g}}\over{\Delta\Omega\Delta\nu}}
\end{equation}
Therefore for the dominant mode the brightness temperature depends only 
on the mode (through $v_{\rm g}$) and on the energy assumed to be available to 
the maser, and not on the saturation length $l_{\rm sat}$. \par
  We estimated the importance of any wave with negative absorption by calculating
a saturation length assuming this particular wave is the only amplified one.
We then compared the various non-dominant saturation lengths
to the saturation length of the dominant mode, where by definition
the dominant mode is the mode with the shortest saturation length.    
    The amplification length of a non-dominant mode is limited to the loss-cone length 
determined by the dominant mode $l_{\rm dom}$. 
The intensity of the non-dominant
mode can then be calculated by using the non-dominant 
absorption and emission coefficients
with the dominant mode`s saturation length. The intensity to be used
in equation \ref{TB} is for this case
\begin{equation}
  I_{\rm non}= {{j_{\rm non}}\over{|k_{\rm non}|}} {\rm e}^{|k_{\rm non}|\cdot l_{\rm dom}}
\end{equation}
or, alternatively
\begin{equation}
I_{\rm non}={{j_{\rm non}}\over{|k_{\rm non}|}}\left({{0.1 E_{\rm k} v_{\rm g} |k_{\rm dom}|}\over
{\Delta\Omega\Delta\nu\ j_{\rm dom}}}\right)^{k_{\rm non}/k_{\rm dom}}
\end{equation}
Where $k_{\rm dom},j_{\rm dom}$ are for the dominant mode, which determines the
saturation length, and $k_{\rm non},j_{\rm non}$ are for some other
mode, at a different frequency and another emission angle, but which
also has a negative absorption, and is therefore amplified. 
This formulation is true no matter which is the dominant mode, also for the
ZM. \par
  If $|k_{\rm non}|\cdot l_{\rm dom}$ is large, the resulting intensity is large
and observable.
  For example for $\nu_{\rm p}/\nu_{\rm B}=1$ (Table \ref{ffp=1}) and $T=5\cdot 10^6 $ K 
the XO saturation length was
the shortest of all modes, though very close to the ZM saturation length.
However, the OM absorption coefficient was also very close
to the XO absorption coefficient, and obviously the non-dominant situation is
relevant. 
This is also one of the rare cases where the OM was slower than the ZM, 
so the growth rate of the OM was much smaller than the growth
rate of the ZM, and certainly smaller than the growth rate of the XO.  
In spite of that, our model predicts a strong emission
in the OM for this table's $\nu_{\rm p}/\nu_{\rm B}=1$. For this case $j/k$ 
for the OM is about $1/16$ of the $j/k$  ratio for the XO.
Therefore even when the power of the exponent is the same, the total intensity
in the OM is only $1/16$ of the intensity in the XO. \par
   If the non-dominant mode has a much smaller absorption, for
example as in $\nu_{\rm p}/\nu_{\rm B}=1$ (Table \ref{ffp=1}) and $T=10^7$ K, 
where $k_{\rm non}\approx k_{\rm dom}/3$,
then the non-dominant intensity is much smaller. For example if 
we assume $k_{\rm dom}\cdot l_{\rm dom} \approx 25-30$, as it often is, and
the same $j/k$ for the dominant and non-dominant modes, 
the factor between the intensities is ${\rm e}^{17}\approx 10^7$.
This factor reduces a brightness temperature of 
$10^{18}$ K to a brightness temperature of only $10^{11}$ K, 
which is lost against the background of the entire loop, and is in any 
case much lower than is observed for microwave spikes. \par
   The way to decide when a non-dominant mode could be observed is to compare
absorption coefficients. In Fig. \ref{growth-rate} the XO growth rate for 
$\nu_{\rm p}/\nu_{\rm B}=1$ (of the example above) appears to be much 
larger than the OM growth rate. However, in Fig. \ref{absorption} we see that the 
absorption coefficients are equal. The example of the previous paragraph shows that 
the OM is observable whenever such a situation occurs.

\subsection{Slow Mode Saturation}
\label{slow-mode}
 Another possibility for observation of a non-dominant mode
is for the case of the dominant mode being slow.
The Z-mode is usually a slow wave, especially near its resonance frequency, 
where its group velocity can be even as low as $c/500$.
It is therefore possible that when 
a certain set of conditions is met another mode is
amplified to its maximum. This can happen if the Z-mode is slow enough
that the other mode reachs its saturation length before the ZM
saturates. In that case until the ZM saturates and reduces
the enegry in the distribution, the other mode is emitted
with the full amplification of its own saturation length. \par
  The growth rate equation \ref{growth} already includes the 
group velocity, and therefore a comparison of growth rates
suffices to show if this slow-mode scenario is possible.
It is also possible to take the group velocity into account from the perspective 
of the spatial
absorption coefficient, though we need a few assumptions.
We assumed a typical emitting electron energy of $\gamma=1.1$, and 
that the important electrons are on the edge of the loss cone with pitch 
angle $\cos(\phi)=0.8$. These parameters gave a velocity parallel to the 
magnetic field of $c/3$ for the most important electrons, while the OM or 
XO wave was usually faster than $c/3$. \par
   For an example we used Table \ref{ffp=1.4} for temperature 
of $10^7$ K. The OM group 
velocity is $0.783\cdot c$, and
the ZM group velocity is $0.1253\cdot c$.
An OM wave of speed $0.783\ c$ beginning near the mirror point at time $t_1$ 
after the first electrons mirrored overtakes these first electrons 
(of parallel velocity $c/3$) at time $t_2=1.74 \cdot t_1$, and at a point 
$0.58\ c\cdot t_1$ from the mirror point. 
In our model (section
\ref{saturation-length}), when this distance is the saturation length,
$0.58\cdot c\cdot t_1=l_{\rm sat}^{OM}$, 
the intensity of the radiation is so great that the end-most volume element 
with a loss-cone distribution is drained of its free energy, and 
the loss-cone distribution disappears from that end-most volume. 
  A Z-mode wave emitted at time $t=0$, and having a group velocity of
$0.1253\ c$, has propagated by this time only to a distance of
$0.1253\cdot c\cdot t_2=0.376\cdot l_{\rm sat}^{OM}$. If the ZM with this group 
velocity has an absorption coefficient such that its saturation length 
is for example $0.4$ of the OM saturation length, the Z-mode had not  
saturated by the time the OM emission has reached its peak and is beginning 
to close the loss-cone.
  However, at some later time the ZM will have had time to saturate near
the mirror-point, and has closed the loss-cone close to the mirror. 
An OM/XO wave starting from the mirror point at that time does not 
see a loss-cone, and 
is not amplified. 
For the case in Table \ref{ffp=1.4} the ZM has a 
saturation length only $0.27$ of the OM saturation length, and
therefore the ZM has saturated by the time the OM wave reaches its 
saturation length. This however does not mean that the OM is not 
amplified as explained below. \par
  A better calculation than the above takes into account that the OM  
wave advances
faster than the electrons and much faster than the ZM wave. 
The requirement then is that the electrons the OM wave passes have
not themselves gone previously through a saturated Z-mode wave, and therefore 
have retained the loss-cone distribution.
  The result of that requirement is that for a mode with $l_{\rm sat}$ to be amplified, 
the Z-mode saturation length $l_{\rm z}$ must be large enough so that the inequality
below is correct :
\begin{equation} 
{{l_{\rm z}}\over{l_{\rm sat}}} \geq {{v_{\rm z}}\over{v_e}} {{v_{\rm g}-v_e}\over{v_{\rm g}-v_{\rm z}}}
\end{equation}
With $v_e$ the velocity of the electrons, $v_{\rm g}$ the group velocity
of the OM or XO, $v_{\rm z}$ the group velocity of the ZM, $l_{\rm z}$ the saturation
length of the ZM, $l_{\rm sat}$ the saturation length of the OM.
 The same equation holds for any two modes, one with group velocity 
$v_{\rm z}$ slower than the electrons, and the other with group velocity 
$v_{\rm g}$ faster than the electrons. \par
   For the case of Table \ref{ffp=1.4}  the left side of the equation is $0.27$
while the right side is $0.257$, and therefore the OM can reach its 
maximum possible amplification before the ZM closes the loss-cone and 
prevents it. \par
   A simpler estimate for the relative importance of the modes is 
to compute $\tau=l_{\rm sat}/v_{\rm g}$ for each of the modes.     
This estimate is very close to calculating the growth rate, since dividing the 
saturation length by $v_{\rm g}$ is equivalent to replacing
the absorption $k$ by the growth rate $\Gamma=-k\cdot v_{\rm g}$ in equation 
\ref{satlen}. This criterion tends to be smaller for a larger growth rate,
and the smallest value is the most important. 
  For our example (Table \ref{ffp=1.4}, $T=10^7$) the 7th column of the table 
shows that the growth rate of the OM is larger than the growth rate of the ZM.
Therefore, even though the saturation length of the OM is much larger than
the saturation length of the ZM, and after the ZM saturates the OM 
disappears, the growth rate criterion, and our calculations above, show
that the OM can be observed for a short duration, until the ZM saturates . \par
  Another example is for $\nu_{\rm p}/\nu_{\rm B}=0.8$ (Table \ref{ffp=0.8}),
$T=5\cdot 10^6$ and the ZM. In the Table we give both the ZM wave 
with largest (absolute value) negative absorption, and the ZM wave with 
the largest growth rate. The ZM wave with the
largest absorption is very slow and has a very small growth rate, and therefore
the important wave is at first the one with the largest growth-rate. 
For $\nu_{\rm p}/\nu_{\rm B}=0.8$ the largest growth rate in the ZM is still smaller than 
the growth  rate of the OM wave, and therefore we conclude that the OM 
reaches its saturation 
length before the ZM closes the loss-cone, and we expect to see an OM 
wave for a short time.
It turns out that the difference between the time to saturation for the OM
and the time to saturation for the largest growth rate ZM is very small, 
about $0.1\ {\rm msec}$. The time
we expect to see the OM at its full amplification is therefore very small. \par
   In this case it was usefull to check the non-dominant-mode approach of the 
previous section, and we found that the OM is observed until the small 
saturation length ZM saturates. When we used the OM absorption coefficient 
with the saturation length of the largest growth rate ZM
we got a brightness temperatures of order $T_{\rm B}=10^{15}$, which
is observable.
 
\subsection{Time Dependence}
  The duration of the pulse of fast electrons is obviously unrelated to the loss-cone 
time scales. 
If the pulse is shorter than the loss-cone time-scale, less than 
$t_2=3\cdot l_{\rm sat}/c$ for the example of the previous section, 
the maser intensity in the OM is not strong enough to close the loss-cone. 
  The loss-cone distribution therefore moves from the 
mirror point at the typical loss-cone velocity, and the maser continues 
to emit, though at a weak intensity. Using the terminology of the example
from the previous section, and assuming a pulse duration of $t_1$,
the maximum amplification length is $c\cdot t_1/3$. In the example we 
had $l_{\rm sat}=0.58\cdot c\cdot t_1$, and therefore the maximum 
amplification length is about $l_{\rm sat}/2$. Since the intensity is exponential with the 
amplification length, this translates into a much weaker emission. 
Typically the power of the exponent for the dominant mode is 20--30, and with 
$l_{\rm sat}/2$ the intensity is more than a hundred thousand times weaker.
For such a case it is unlikely that the maser emission can be observed at 
all. \par
  The other extreme is that the pulse of fast electrons persists for a time 
much longer than typical maser saturation times. For such a case, the loss-cone 
closes at distance $l_{\rm sat}$ from the mirror point, but new fast electrons 
move into place constantly. At first glance it would seem that the maser 
persists for approximately the duration of the pulse, however, if there
is a ZM with a smaller saturation length than the observable modes this
does not 
happen. After the Z-mode reaches its saturation length it quenches the fast 
electron loss-cone distribution, and the observable modes can not be amplified
enough to be seen. 
In this picture the ZM waves saturate and maintain an intensity just strong 
enough to drain every new volume element of its energy when it 
overtakes the ZM waves.  An observable OM or XO wave starting from the mirror 
at a time $t > t_1$ of the previous section, so that the ZM has already reached 
saturation $t > l_{\rm sat}^{ZM}/v_{\rm g}^{ZM}$, does not see a loss-cone, and therefore is 
not amplified.
For our example above, the intensity of an observable mode at first increases, reaching
a maximum at time $t_2$ after the first electrons are mirrored. At this time  
the OM wave which left the mirror at $t_1$ reaches $l_{\rm sat}$. This
high intensity can last only until OM waves leaving the mirror encounter
saturated ZM waves. For the example from Table \ref{ffp=1.4} the ZM 
reaches its saturation length near the mirror at time $1.25\cdot t_1$. 
An OM wave leaving the mirror point at $1.25\cdot t_1$
reaches $l_{\rm sat}$ at $t_2+0.25 t_1$, and the maser then starts to decrease
in intensity. The maser is therefore at peak intensity only for
a time  $0.25 t_1=1.7$ msec.  
Of course, if the shortest saturation length is for an observable mode,
the maser does persist for the length of the pulse of accelerated particles. \par
   Our conclusion is therefore that if the smallest $l_{\rm sat}$ is for the ZM, the 
maser persists for a time less then but of order $[l_{\rm sat}/v_{\rm g}]_{ZM}$, regardless 
of the duration of the fast electron acceleration pulse. This is the case
for Table \ref{ffp=0.8} at all temperatures. Of course, after the pulse
ends, the waves disappear, and when a new pulse reaches the mirror
point, the process can start again. \par
  An example of a different kind is for $\nu_{\rm p}/\nu_{\rm B}=1$ (Table \ref{ffp=1})
and $T=5\cdot 10^6$ K. For this case the saturation length
of the XO was shorter than the saturation length of the ZM, and the
saturation length of the OM was not much longer. Therefore we expect
the XO radiation to last, and the OM radiation to be amplified as well.
For the other tables, at $T=5\cdot 10^6$ the smallest saturation
length was always for an observable mode, and therefore at this temperature 
the loss cone should always produce long lasting maser emission. \par  
  The differential approach described in section \ref{saturation-length} also gives a 
model for the time-profile of 
the maser. The emission should appear as a rising exponent, and 
then decrease faster than an exponent, because the elements whose 
energy was exhausted begin to positively absorb radiation 
from the still amplifying elements.   
  This picture is at odds with the characteristics observed, 
which were a rise as ${\rm e}^(t^2)$ and a decrease as an exponent 
\cite{G+B}.
Perhaps the gaussian rise is a feature of the acceleration
mechanism, for example if  $n_{\rm hot}$ also increases with time. \par

\section{Comparison with Previous Work}
  We can compare our results with only one previous case
 \cite{Aschwanden b} 
where saturation lengths were estimated, and given for example for 
$\nu_{\rm p}/\nu_{\rm B}=0.5$ as 
$l_{\rm sat}^{ZM}=10$ km and
$l_{\rm sat}^{OM}=30$ km.\par
  We incorporated in our program Aschwanden`s thermal distribution in energy
and $\sin^6(3\times\phi)$ loss-cone. Aschwanden used the standard parameters
$T_{\rm hot}=10^8$ K, $T_{\rm cold}=10^6$ K, $n_{\rm cold}=1.25\cdot 10^8\ {\rm cm^{-3}}$,
and $n_{\rm hot}/n_{\rm cold}=0.01$. 
With the above and from the standard $\nu_{\rm p}/\nu_{\rm B}=0.1$ a magnetic field 
of $B=360$ gauss was the standard field. 
If the number of particles was kept constant
the magnetic field for $\nu_{\rm p}/\nu_{\rm B}=0.5$ as in the example above should
have been reduced to $B=72$ gauss. \par
  Our calculations with Aschwanden`s distribution functions gave 
the normalized results, with $B=100$ gauss and divided by $n_{\rm hot}$, 
for the OM:
$k=-1.1\cdot 10^{-11}$, $j=1.69\cdot 10^{-24}$, $c/v_{\rm g}=1.144$,
$\nu/\nu_{\rm B}=1.019$, $\cos(\theta)=0.24$
and for the ZM:
$k=-3.6\cdot 10^{-11}$, $j=4\cdot 10^{-23}$, $c/v_{\rm g}=7.3$,
$\nu/\nu_{\rm B}=1.039$, $\cos(\theta)=0.2$.
Where these numbers are for the respective largest growth rate
in every mode, since that was Aschwanden`s criterion for the dominant 
wave. \par
  Using our simple method to calculate the saturation length,
and the scaling laws $k\propto n_{\rm hot}/B$ and $j \propto n_{\rm hot}\times B$,
we got
$l_{\rm sat}^{ZM}=4.89$ km and
$l_{\rm sat}^{OM}=17.9$ km.
These saturation lengths are about a factor $2$ smaller than Aschwanden`s.
We also have to take into account Aschwanden`s assumption of an initial 
energy density equivalent to $T_{\rm B}=10^{14}$ K. In our model the waves amplify 
the spontaneous emission, and the amplification is according to the radiative 
transfer equation \ref{radtran}.
A typical value is $j/k \approx 10^{-13}$ and if this is the intensity $I$ 
the equation for the brightness temperature (equation \ref{TB}) gives the equivalent 
temperature as only $T_{\rm B}=10^7$ K. The brightness temperature reaches 
$T_{\rm B}=10^{14}$ K only after the wave has been amplified by a factor of 
$10^7 \approx {\rm e}^{16}$, and the comparison of the saturation lengths
should be only from this stage. \par
  The total saturation length gives a power in the exponent of about $30$, 
therefore if we had started with a wave at $T_{\rm B}=10^{14}$ K the saturation
length would have been only half of what we calculate. 
The conclusion is that computing the saturation length with our method
resulted in saturation lengths of a factor of 4 smaller than those calculated by 
Aschwanden`s method.  This was the result of neglecting the change in the
loss-cone during the amplification of the wave, and assuming that 
the absorption coefficient remains constant until the wave reaches 
its maximum energy. The differential iterative method shows that
the actual saturation length is up to twice what the simple
method gives, and therefore we get a factor of about $2$ back.
Considering the simplicity of the equations we use, a factor of two
is extremely close.

\section{Summary}
   We have developed a model for estimating the intensity and duration
of emission through the Electron-Cyclotron-Maser (ECM) mechanism. \par
   Observations led us to use a power law for the fast electrons, with a 
minimum energy of $10$ keV and maximum energy of $1$ MeV. 
This distribution is harder than the ones studied in the past 
(Winglee \& Dulk 1986; Aschwanden 1990a). \par
   From recent results \cite{B+A} we estimate the 
important ratio $\nu_{\rm p}/\nu_{\rm B}$ in the foot-points of flares, which is the region 
of interest for this study, to be in the range 0.5--3.
    For most of this range a loss cone in a power law distribution of electrons results 
in the ZM being the dominant magneto-ionic mode. The ZM is unobservable, because  
when a ZM wave reaches a region with a smaller plasma frequency 
than the plasma frequency where the wave originated,
it is 
absorbed. Thus it seems at first glance that the maser mechanism can not produce 
observable emission. \par
   We use an absorption coefficient approach, differing from previous
studies which used the growth rate. Using this approach we calculate
the {\it saturation length} for the competing magneto-ionic modes, and show
that in many cases the observable OM and XO are amplified to
observable intensities. \par
  The conditions which enable the OM and XO modes to be amplified
in the presence of a ZM with much larger (in absolute magnitude) 
negative absorption are : first, the inclusion of an ambient thermal
plasma with temperature of order $T=10^7$ K ; second, the low
group velocity of the ZM which enables a competing mode to reach
its saturation before the ZM saturates ; third, the high amplification
of the maser, which makes even non-dominant modes observable. \par
  Our conclusion is that in the range $\nu_{\rm p}/\nu_{\rm B}=0.5-1.8$ the maser
mechanism can produce observable emission in the form of intense
spikes with a narrow bandwidth -- about one per cent of the frequency -- and narrow
angular width -- about $0.1 \pi$ solid angle. \par
   The examples given in the tables indicate that long duration emission
in the OM and XO appears at angles of 30--60 degrees to the magnetic field, 
and at 3--6 per cent of the cyclotron frequency above the nearest 
cyclotron harmonic. This is usually the second harmonic for the OM,
and always the second harmonic for the XO.
We expect the long duration emission to be determined by the duration
of the pulse of accelerated electrons, and the brightness temperature
to be of order $T_{\rm B}=10^{16}$ K and above. \par
  Short duration emission appears when the OM or XO are emitted
only until the ZM saturates. For these cases the saturation lengths of the OM
and XO are longer than the ZM saturation length, but the ZM group velocity
is smaller than the group velocity of the OM or XO. The short duration
emission lasts in our model from fractions of a millisecond to a millisecond,
for example for $\nu_{\rm p}/\nu_{\rm B}=0.8$ (Table \ref{ffp=0.8}) and a temperature 
of $5\cdot 10^6$ K, where the OM has the largest growth rate. \par
  From the results given in Tables \ref{ffp=1} to  \ref{ffp=1.8} for a temperature 
of $5\cdot 10^6$ K, we expect emission for the parameters of the tables 
for the duration of the accelerated electrons` pulse, and at frequency 
$\nu \approx 2.06\nu_{\rm B}$. The exception is the case of $\nu_{\rm p}/\nu_{\rm B}=1$
(Table \ref{ffp=1}) where the XO emission for $T=5\cdot 10^6$ K 
is at $\nu/\nu_{\rm B} \approx 2.06$, and there is also OM emission of observable 
intensity at $\nu/\nu_{\rm B} \approx 1.03$. 
  We also expect long duration emission for $\nu_{\rm p}/\nu_{\rm B}=1.2$
(Table \ref{ffp=1.2}) and $T=10^7$ K. For this case 
$l_{\rm sat}^{XO} \approx l_{\rm sat}^{ZM}$, 
and therefore the XO will be amplified to observable intensities.
The XO frequency is $\nu = 2.129\nu_{\rm B}$.  
  For the cases $\nu_{\rm p}/\nu_{\rm B}=1.6$ and $\nu_{\rm p}/\nu_{\rm B}=1.8$ (Table \ref{ffp=1.6} and Table \ref{ffp=1.8}) and $T=10^7$ and 
$T=2\cdot 10^7$, the OM is the dominant mode, but the saturation lengths
are much too large. We conclude that the loss-cone relaxes
by some other means, and not through the ECM mechanism. \par
  In summary, this study shows that the ECM produces millisecond radio
spikes such as observed from the Sun. The ECM emission reproduces the frequency
of the observed spikes, the bandwidth of the spikes, the source size as recently
observed \cite{Benzetal}, the brightness temperature,
and the duration of the spikes. \par
  The ECM does not produce emission at two
frequencies simultaneously, and therefore can not reproduce the observed bands \cite{Krucker}. However, if a loss cone distribution
forms independently in different regions, and the magnetic field is not the same in these 
regions, then ECM in several frequencies and angles can be produced. For example
in Table \ref{ffp=1} for temperature $T=5\cdot 10^6$ the XO emission is at
angle $69$ degrees. In Table \ref{ffp=1.2} for $T=5\cdot 10^6$ the XO emission
is at angle $64$ degrees. If two regions, one with $\nu_{\rm p}/\nu_{\rm B}=1$ and one with
$\nu_{\rm p}/\nu_{\rm B}=1.2$, emit ECM, and if the magnetic field changes direction by a few degrees
between the regions, it may become possible to observe both. Since the magnetic field
and the density both increase towards the foot points, the ratio between the emitted frequencies
in the two regions can be larger than $1.2$. \par
  Our model also does not reproduce the reported gaussian rise time \cite{G+B}. 
However, if the number of fast electrons also increases with time,the rise time 
is a feature of the acceleration mechanism, since the emission in ECM is directly
proportional to the number density of the fast electrons. \par  
 It is possible to conclude from our results that the observed frequency 
of the spikes should be higher at the beginning of a flare, since the hot and dense 
chromospheric plasma is then still close to the chromosphere, where a large 
magnetic field causes a large cyclotron frequency. As the flare advances,
and the evaporation front propagates upwards, temperatures below increase, 
the unobservable ZM dominates for the larger temperatures, and the magnetic 
field in the region from which ECM emission is observed is smaller. \par

\clearpage

\begin{table}
\begin{minipage}{80mm}
\caption {$\nu_{\rm p}/\nu_{\rm B}=0.8$\label{ffp=0.8}}
\begin{tabular}{ c c c c c c c c }

$T/10^6$ & $\nu/\nu_{\rm B}$ & emission angle & $k_{\rm max}$ & $l_{\rm sat}/10^6$ & $v_{\rm g}$ & $\Gamma$ & Mode \\
Kelvin&  & Degrees &${\rm cm^2}$& cm & ${\rm cm s^{-1}}$ & ${\rm cm^3 s^{-1}}$ & \\
\hline
5 &1.2745&82&-2.63E-10&0.064883&2.05E+07&5.37E-03&ZM\\
5 &1.0351&75&-1.8E-11&1.947&1.13E+10&0.204&ZM\\
5 &1.03&64&-1.34E-11&2.788&1.89E+10&0.253&OM\\
5 &2.0616&70&-7.67E-12&4.57&2.4E+10&0.184&XO\\
10 &1.2709&81&-6.68E-11&0.28&5.25E+07&3.5E-03&ZM\\
10 &1.087&69&-1.01E-11&3.26&7.89E+09&0.08&ZM\\
10 &1.079&52&-1.71E-12&21&2.11E+10&0.036&OM\\
10 &2.131&61&-1.12E-12&29.7&2.45E+10&0.027&XO\\
20 &1.2663&80&-4.23E-11&0.519&1.08E+08&4.6E-03&ZM\\
20 &1.2121&70&-1.54E-11&1.801&1.38E+09&0.02&ZM\\
20 &1.189&37&-8.48E-14&400.9&2.52E+10&2.1E-03&OM\\
\hline
\end{tabular}
\end{minipage}
\end{table}

\begin{table}
\begin{minipage}{80mm}
\caption {$\nu_{\rm p}/\nu_{\rm B}=1$\label{ffp=1}}
\begin{tabular}{ c c c c c c c c }

$T/10^6$ & $\nu/\nu_{\rm B}$ & emission angle & $k_{\rm max}$ & $l_{\rm sat}/10^6$ & $v_{\rm g}$ & $\Gamma$ & Mode \\
Kelvin&  & Degrees &${\rm cm^2}$& cm & ${\rm cm s^{-1}}$ & 
${\rm cm^3 s^{-1}}$ & \\
\hline
5&1.036&72&-7.76E-12&2.99&1.35E+10&0.105&ZM\\
5 &1.0301&38&-7.66E-12&3.16&5.14E+09&0.039&OM\\
5 &2.057&69&-7.75E-12&2.98&1.99E+10&0.154&XO\\
10&1.0909&64&-4.05E-12&5.44&1.1E+10&0.044&ZM\\
10&1.076&35&-1.39E-12&16.54&9.86E+09&0.014&OM\\
20&1.3265&66&-3.9E-12&4.34&1.4E+09&5.5E-03&ZM\\
20&1.1872&26&-4.47E-14&492.6&2.09E+10&9.37E-04&OM\\
\hline
\end{tabular}
\end{minipage}
\end{table}

\begin{table}
\begin{minipage}{80mm}
\caption {$\nu_{\rm p}/\nu_{\rm B}=1.2$\label{ffp=1.2}}
\begin{tabular}{ c c c c c c c c }

$T/10^6$ & $\nu/\nu_{\rm B}$ & emission angle & $k_{\rm max}$ & $l_{\rm sat}/10^6$ & $v_{\rm g}$ &
$\Gamma$ & Mode \\
Kelvin&  & Degrees &${\rm cm^2}$& cm & ${\rm cm s^{-1}}$ & ${\rm cm^3 s^{-1}}$ & \\
\hline
5&1.208&47&-1E-12&14.7&7.89E+09&7.9E-03&ZM\\
5&2.061&69&-5.7E-13&28.5&2.48E+10&0.014&OM\\
5&2.06&64&-7.9E-12&2.05&1.4E+10&0.111&XO\\
5&1.9653&89&-4.2E-12&3.9&7.8E+09&0.0323&XO\\
10&1.286&50&-1.08E-12&12.8&5.8E+09&6.2E-03&ZM\\
10&2.129&52&-1.08E-12&14.33&1.57E+10&0.017&XO\\
20&1.286&50&-1.08E-12&12.8&5.8E+09 &6.2E-03 & ZM\\
20&2.319&33&-7.65E-14&190&1.9E+10&1.45E-03&XO\\
\hline
\end{tabular}
\end{minipage}
\end{table}

\begin{table}
\begin{minipage}{80mm}
\caption {$\nu_{\rm p}/\nu_{\rm B}=1.4$\label{ffp=1.4}}
\begin{tabular}{ c c c c c c c c }

$T/10^6$ & $\nu/\nu_{\rm B}$ & emission angle & $k_{\rm max}$ & $l_{\rm sat}/10^6$ & $v_{\rm g}$ &
$\Gamma$ & Mode \\
Kelvin&  & Degrees &${\rm cm^2}$& cm & ${\rm cm s^{-1}}$ & ${\rm cm^3 s^{-1}}$ & \\
\hline
5&1.422&37&-3E-13&32.3&3.76E+9&1.1E-03&ZM\\
5&2.0597&67&-4.3E-13&27.9&2.2E+10&9.7E-03&OM\\
5&2.056&31&-1.5E-12&8.5&6.4E+9&9.3E-03&XO\\
10&1.42&37&-3E-13&32.3&3.76E+9&1.1E-03&ZM\\
10&2.13&58&-9.7E-14&119.8&2.35E+10&2.3E-03&OM\\
20&1.42&37&-3E-13&32.3&3.76E+9&1.1E-03&ZM\\
\hline
\end{tabular}
\end{minipage}
\end{table}

\begin{table}
\begin{minipage}{80mm}
\caption {$\nu_{\rm p}/\nu_{\rm B}=1.6$\label{ffp=1.6}}
\begin{tabular}{ c c c c c c c c }

$T/10^6$ & $\nu/\nu_{\rm B}$ & emission angle & $k_{\rm max}$ & $l_{\rm sat}/10^6$ & $v_{\rm g}$ &
$\Gamma$ & Mode \\
Kelvin&  & Degrees &${\rm cm^2}$& cm & ${\rm cm s^{-1}}$ & ${\rm cm^3 s^{-1}}$ & \\
\hline
5&2.0589&64&-2.91E-13&32.3&1.91E+10&5.6E-03&OM\\
5&3.0889&69&-6.4E-14&129.4&2.28E+10&1.5E-03&XO\\
10&2.123&55&-6.72E-14&133.96&2.08E+10&1.4E-03&OM\\
20&2.285&41&-4.84E-15&1761.7&2.38E+10&1.1E-04&OM\\
\hline
\end{tabular}
\end{minipage}
\end{table}

\clearpage

\begin{table}
\begin{minipage}{80mm}
\caption {$\nu_{\rm p}/\nu_{\rm B}=1.8$\label{ffp=1.8}}
\begin{tabular}{ c c c c c c c c }

$T/10^6$ & $\nu/\nu_{\rm B}$ & emission angle & $k_{\rm max}$ & $l_{\rm sat}/10^6$ & $v_{\rm g}$ &
$\Gamma$ & Mode \\
Kelvin&  & Degrees &${\rm cm^2}$& cm & ${\rm cm s^{-1}}$ & ${\rm cm^3 s^{-1}}$ & \\
\hline
5&2.04885&86.6&-2.65E-08&0.00019&6.45E+07&1.714&ZMN\\
5&2.0596&57&-1.5E-13&48.9&1.44E+10&2.2E-03&OM\\
5&3.0844&68&-5.4E-14&122.3&2.3E+10&1.1E-03&XO\\
10&2.1237&48&-3.97E-14&180.8&1.68E+10&6.7E-04&OM\\
10&3.1539&60&-1.94E-14&331.3&2.08E+10&4.03E-04&XO\\
20&2.2893&34&-2.24E-15&3045&2.18E+10&4.9E-05&OM\\
\hline
\end{tabular}
\end{minipage}
\end{table}

\clearpage

\clearpage

\begin{thebibliography}{12345678901234567890}
\bibitem[\protect\citename {Aschwanden} 1990a]{Aschwanden a} Aschwanden M.J,  1990a, A\&AS, 85, 1141
\bibitem[\protect\citename {Aschwanden} 1990b]{Aschwanden b}  Aschwanden M.J.,  1990b, A\&A, 237,512
\bibitem[\protect\citename {Aschwanden \& Benz} 1988]{B+A2}  Aschwanden M.J., Benz A.O. , 1988, A\&A, 332,447
\bibitem[\protect\citename {Aschwanden \& Benz} 1997]{B+A} Aschwanden M.J., Benz A.O. ,  1997, ApJ, 480,825
\bibitem[\protect\citename {Benz} 1986]{Benz} Benz  A.O.,  1986,Sol.~Phys., 104,  99
\bibitem[\protect\citename {Benz} 1993]{BenzBook} Benz A.O., 1993, Plasma Astrophysics Kinetic Processes in Solar and Stellar Coronae. Kluwer Academic Publishers
\bibitem[\protect\citename {Benz et.al.} 1996]{Benzetal} Benz A.O., Graham D., Isliker H., Andersson C., Koehnlein W.,  Mantovani F.,  Umana G.,  1996, A\&A 305, 970
\bibitem[\protect\citename {Fleishman \& Yastrebov} 1994a]{Fleishman1} Fleishman G.D., Yastrebov S.G.,  1994a, Sol.~Phys., 154, 361
\bibitem[\protect\citename {Fleishman \& Yastrebov} 1994b]{Fleishman2} Fleishman G.D.,  Yastrebov S.G.,  1994b, Sol.~Phys., 153, 389
\bibitem[\protect\citename {Gudel \& Benz} 1990]{G+B} Gudel M., Benz A.O.,  1990,A\&A, 231,202
\bibitem[\protect\citename {Holman \& Benka} 1992]{HandB} Holman G.D., Benka S.G.,  1992,ApJ, 391, 854
\bibitem[\protect\citename {Holman, Eichler \& Kundu} 1980]{Holman} Holman G.D., Eichler D., Kundu M.R., 1980, in M.R. Kundu, T.E. Gergely , eds, Proc. IAU Symp. 86, Radio Physics of the Sun. Dordrecht : Reidel, p. 457
\bibitem[\protect\citename {Kuncic \& Robinson} 1993]{Kuncic} Kuncic Z., Robinson P.A.,  1993,Sol.~Phys., 145,317
\bibitem[\protect\citename {Krucker \& Benz} 1994]{Krucker} Krucker S.,  Benz A.O.,   1994,A\&A, 285,1038
\bibitem[\protect\citename {Lang} 1980]{Lang} Lang K.R., 1980, Astrophysical Formulae 2nd Edition. Springer Verlag
\bibitem[\protect\citename {Mackinnon, Vlahos \&Vilmer} 1992]{Mackinnon}  Mackinnon A. , Vlahos L., Vilmer N., 1992,A\&A, 256, 2, 613
\bibitem[\protect\citename {Melrose} 1989]{MelroseB} Melrose D.B, 1989, Instabilities in Space and Laboratory Plasma.  Cambridge University Press
\bibitem[\protect\citename {Melrose \& Dulk} 1982]{D+M} Melrose D.B, Dulk G. A.,  1982, ApJ, 259, 844
\bibitem[\protect\citename {Ramaty} 1969]{R} Ramaty R., 1969, ApJ , 158, 753
\bibitem[\protect\citename {Trulsen \& Fejer} 1970]{Trulsen} Trulsen J., Fejer J.A.,  1970, J. Plas. Phys., 4,4, 825
\bibitem[\protect\citename {White, Melrose \& Dulk} 1986]{WMD} White S.M, Melrose D.B., Dulk G.A., 1986, Adv. Space Res. 6,6,163
\bibitem[\protect\citename {Winglee} 1985]{Winglee} Winglee R.M.,1985, ApJ, 291,160
\bibitem[\protect\citename {Winglee \& Dulk} 1986]{Winglee2} Winglee R.M, Dulk G.A.,  1986,Sol.~Phys., 104,93
\bibitem[\protect\citename {Wu \& Lee} 1979]{Wu} Wu C.S., Lee L.C, 1979, ApJ, 230,621
\bibitem[\protect\citename {Zirin} 1989]{Zirin}Zirin H., 1989, Astrophysics of the Sun. Cambridge University Press
\end{thebibliography}
\end{document}